\documentclass[12pt]{article}
\usepackage{graphicx,amssymb,amsmath}

\newcommand{\be}{\begin{equation}}
\newcommand{\ee}{\end{equation}}
\newcommand{\bea}{\begin{eqnarray}}
\newcommand{\eea}{\end{eqnarray}}
\newcommand{\beas}{\begin{eqnarray*}}
\newcommand{\eeas}{\end{eqnarray*}}

\newcommand{\DS}{{\Delta_{\textstyle *}}}

\begin{document}
\begin{titlepage}

\begin{center}

{\Large Bulk equations of motion from CFT correlators}

\vspace{8mm}

\renewcommand\thefootnote{\mbox{$\fnsymbol{footnote}$}}
Daniel Kabat${}^{1}$\footnote{daniel.kabat@lehman.cuny.edu} and
Gilad Lifschytz${}^{1,2,3}$\footnote{giladl@research.haifa.ac.il}

\vspace{4mm}

${}^1${\small \sl Department of Physics and Astronomy} \\
{\small \sl Lehman College, City University of New York, Bronx NY 10468, USA}

\vspace{2mm}

${}^2${\small \sl Physics Department} \\
{\small \sl City College, City University of New York, New York NY 10031, USA}

\vspace{2mm}
${}^3${\small \sl Department of Mathematics and Physics} \\
{\small \sl University of Haifa at Oranim, Kiryat Tivon 36006, Israel}

\end{center}

\vspace{8mm}

\noindent
To ${\cal O}(1/N)$ we derive, purely from CFT data, the bulk equations
of motion for interacting scalar fields and for scalars coupled to
gauge fields and gravity.  We first uplift CFT operators to mimic
local AdS fields by imposing bulk microcausality.  This requires
adding an infinite tower of smeared higher-dimension double-trace
operators to the CFT definition of a bulk field, with coefficients
that we explicitly compute.  By summing the contribution of the
higher-dimension operators we derive the equations of motion satisfied
by these uplifted CFT operators and show that we precisely recover the
expected bulk equations of motion.  We exhibit the freedom in the CFT
construction which corresponds to bulk field redefinitions.

\end{titlepage}

\setcounter{footnote}{0}
\renewcommand\thefootnote{\mbox{\arabic{footnote}}}

\section{Introduction and overview\label{sect:intro}}
The AdS/CFT  correspondence \cite{Maldacena:1997re} states that AdS quantum gravity is holographically encoded in the boundary CFT.  To extract bulk properties we must be able to decode the hologram, that is, we must be able to reconstruct the bulk physics of interest from CFT data.  A particularly straightforward way to recover bulk physics is to express local fields in AdS in terms of the CFT.  This project has been pursued since the
early days of AdS/CFT.  At first free scalar fields were considered \cite{Balasubramanian:1998sn,Banks:1998dd,
Dobrev:1998md,Bena:1999jv,Hamilton:2005ju,Hamilton:2006az,Hamilton:2006fh}, but later the program was extended to include interactions
\cite{Kabat:2011rz,Heemskerk:2012mn} and fields with spin \cite{Heemskerk:2012np,Kabat:2012hp,Kabat:2012av,Kabat:2013wga,Sarkar:2014dma}.

In the literature two approaches to constructing local bulk fields have been discussed.
\begin{itemize}
\item
One can solve the bulk equations of motion perturbatively, to determine the radial profile of bulk fields in terms of boundary conditions that match on to the CFT.  This approach has the advantage that it more-or-less manifestly reproduces perturbative effective field theory in the bulk.  The disadvantage of this approach is that it requires knowing the bulk equations of motion.
\item
One can start from CFT correlators and attempt to lift them to define correlation functions in the bulk, in a way which respects bulk locality.  This requires
redefining the bulk field to include an infinite tower of higher-dimension operators.  The advantage of this approach is that it requires very little input from
the bulk.\footnote{It relies on the map from CFT operators to free fields in the bulk, which can be understood from representation theory \cite{Dobrev:1998md}.
Also for fields with spin it relies on understanding the appropriate notion of bulk microcausality \cite{Kabat:2013wga}.}  A disadvantage is that it was not so clear this approach
can be carried out explicitly, nor was it obvious it will agree with bulk effective field theory.
\end{itemize}
The purpose of this paper is to show that the program of implementing bulk locality can be carried out to ${\cal O}(1/N)$ for interacting scalar fields as well as for scalars coupled
to gauge fields and gravity.  To this order in $1/N$, that is at the level of three-point functions, we show that it reproduces effective field theory in the bulk (meaning it exactly agrees with the first approach) and in fact lets us
determine the bulk equations of motion starting from CFT correlators.  This extends our previous work by showing that the coefficients of all higher-dimension operators can
be explicitly determined, and that the entire tower of higher-dimension operators can be re-summed to reproduce the expected local physics in the bulk.

\bigskip

Let us summarize the approach to constructing bulk fields based on locality.  For more details see \cite{Kabat:2011rz,Kabat:2012av,Kabat:2013wga}.
Throughout this paper we consider AdS${}_{d+1}$ / CFT${}_d$ in Poincar\'e coordinates, with metric
\[
ds^2 = {R^2 \over z^2} \left(-dt^2 + \vert d\vec{x} \vert^2 + dz^2 \right)
\]
We'll set the AdS radius of curvature $R = 1$.  The boundary coordinates are collectively denoted by $x^\mu = (t,\vec{x})$, and the full set of AdS coordinates are denoted
$X^M = (x^\mu,z)$.

We assume the CFT has a matrix $1/N$ expansion. To leading order in $1/N$, i.e.\ at the level of two-point functions, we can define a CFT operator
$\phi_i(x,z)$ whose two-point function will mimic the bulk two-point function by setting
\begin{equation}
\phi_i(x,z)=\int d^d x' K_i(x,z \vert x'){\cal O}_i(x').
\end{equation}
Here ${\cal O}_i$ is a primary scalar operator of dimension $\Delta_i$ in the CFT, and the smearing function $K_i$ obeys a free wave equation in the bulk.
\be
\label{FreeSmear}
\left(\nabla^2 - m_i^2\right) K_i = 0 \qquad\qquad m_i^2 = \Delta_i(\Delta_i - d)
\ee

At ${\cal O}(1/N)$ we need to take into account the CFT three-point function of primary single-trace operators $\langle {\cal O}_i(x) {\cal O}_j(y_1) {\cal O}_k(y_2) \rangle$.  As above we'd like to lift the first operator ${\cal O}_i(x)$ to behave like a local field in the bulk $\phi_i(x,z)$.  As a first attempt one could simply
integrate the CFT correlator against the appropriate smearing function $K_i$.
\be
\langle \phi_i(x,z) {\cal O}_j(y_1) {\cal O}_k(y_2) \rangle \mathop=^? \int d^d x' K_i(x,z \vert x') \langle {\cal O}_i(x') {\cal O}_j(y_1) {\cal O}_k(y_2) \rangle
\label{naive}
\ee
There is, of course, something odd about this construction.  The smearing function $K_i$ obeys a free wave equation in the bulk.
So it would appear we've constructed a free field in the bulk, even though the CFT has a non-trivial 3-point interaction.

To understand what's gone wrong in (\ref{naive}) it's useful to evaluate the smearing integral explicitly.  It's convenient to work in terms of
\be
\label{chidef}
\chi=\frac{[(x-y_1)^2+z^2][(x-y_2)^2+z^2]}{(y_1-y_2)^2 z^2}
\ee
which is an AdS-invariant cross-ratio depending on one bulk point and two boundary points.
Note that $\chi$ vanishes when the bulk point is light-like separated from either of the boundary points and diverges when the two boundary points are null separated.  When all three points are spacelike separated we have $\chi > 0$.\footnote{A helpful limit to keep in mind is $y_2$ $\rightarrow$ spatial
infinity, which sets $\chi = {(x - y_1)^2 + z^2 \over z^2}$.  In this limit note that $\chi = 1$ corresponds to $x$ and $y_1$ being null separated on the boundary.
Away from this limit $\chi = 1$ does not have such a simple geometric interpretation, although it always corresponds to spacelike separation in the bulk.}

The smearing integral was evaluated in \cite{Kabat:2011rz,Kabat:2012av}, where it was found that
\be
\label{NaiveSmear}
\int d^d x' K_i(x,z \vert x') \langle {\cal O}_i(x') {\cal O}_j(y_1) {\cal O}_k(y_2) \rangle =\frac{1}{(y_1 - y_2)^{2\Delta_j}}
\left[\frac{z}{z^2+(x-y_2)^2}\right]^{\Delta_k-\Delta_j} I_{ijk}(\chi)
\ee
The explicit form of the function $I_{ijk}(\chi)$ is given below in (\ref{Idef}).  We find that $I_{ijk}$ is singular
at $\chi = 1$ and in general has a cut for $0 < \chi < 1$.
Now we can see what's wrong with the naive expression (\ref{NaiveSmear}).  The commutator $\big\langle [\phi_i,{\cal O}_j] {\cal O}_k \big\rangle$ is given by the discontinuity across the cut of $I_{ijk}$.\footnote{The bulk Wightman function $\langle \phi_i(x,z) {\cal O}_j(y_1) {\cal O}_k(y_2) \rangle$ is defined with a $t \rightarrow t - i \epsilon$
prescription, or equivalently $\chi \rightarrow \chi + i \epsilon \, {\rm sign}(t - t_1)$.  Likewise $\langle {\cal O}_j(y_1) \phi_i(x,z) {\cal O}_k(y_2)\rangle$ is defined with $t \rightarrow t + i \epsilon$, or $\chi \rightarrow \chi - i \epsilon \, {\rm sign}(t - t_1)$.  Thus the commutator
picks up the discontinuity across the cut.}  Since $\chi = 1$ corresponds to spacelike separation, this means the bulk field constructed in (\ref{NaiveSmear}) will not
commute with boundary operators, even though they're spacelike separated.  The naive expression (\ref{NaiveSmear}) indeed produces a free field in the bulk,
however it's a field which violates bulk locality.

To restore bulk locality, the idea presented in \cite{Kabat:2011rz} and elaborated in \cite{Kabat:2012av,Kabat:2013wga} is to modify the definition of the bulk scalar
field by adding a tower of smeared higher-dimension primary scalar operators ${\cal O}_n$.  To ${\cal O}(1/N)$, the operators ${\cal O}_n$ will be double-trace
operators built from a product of ${\cal O}_j$ and ${\cal O}_k$ with $2n$ derivatives.  These operators have non-trivial 3-point functions with ${\cal O}_j$ and ${\cal O}_k$, and by choosing their coefficients appropriately, it should be possible to cancel the unwanted non-analyticity in the correlator.
So the proposal was that a local bulk field can be built by setting\footnote{The smearing functions $K_n$ are constructed, just as in (\ref{FreeSmear}), to satisfy a free wave equation
in the bulk, $(\nabla^2 - m_n^2) K_n = 0$ with $m_n^2 = \Delta_n (\Delta_n - d)$.  Note that $K_n$ only depends on the mass of the bulk field or equivalently the conformal dimension
of the boundary operator.  The fact that the ${\cal O}_n$ are double-trace doesn't affect the smearing function.}
\begin{equation}
\phi_{i}(x,z)=\int d^dx' K_{i}(x,z \vert x') {\cal O}_i(x') + {1 \over N} \sum_n a^{CFT}_{n} \int d^dx' K_{n}(x,z \vert x') {\cal O}_{n}(x')
\label{basicbulkcft}
\end{equation}
where the coefficients $a_{n}^{CFT}$ are chosen to give a 3-point correlator (computed purely in the CFT) that is analytic for $\chi>0$.\footnote{This procedure will restore locality with a specific pair of boundary
operators ${\cal O}_j$ and ${\cal O}_k$.  To ${\cal O}(1/N)$, to restore locality with all pairs simultaneously, one can simply superimpose the towers of higher-dimension operators.}

In this paper we will give some explicit examples of this construction.  We will show that all of the coefficients $a_n^{CFT}$ can be explicitly determined by requiring bulk locality, and that the resulting expression for a bulk field (\ref{basicbulkcft}) in fact obeys the expected bulk equation of motion. We also exhibit the freedom in the CFT construction which corresponds to bulk field redefinitions. Since the calculations which follow are rather lengthy, we
provide here a brief summary of our main results.

\noindent{\em Interacting scalar fields}\\
\noindent
We treat a number of examples of interacting scalar fields explicitly.  In section \ref{sect:d=1} we start off with massless scalar fields in AdS${}_2$, corresponding to operators
with dimension $\Delta = 1$, and we go on to consider an operator of dimension $\Delta_\nu$ interacting with two operators of dimension $\Delta_j = \Delta_k = l$.
These AdS${}_2$ results are extended to scalars in AdS${}_3$ and AdS${}_4$ in sections \ref{sect:d=2} and \ref{sect:d=3}.  In all cases we find that the coefficients of
the higher-dimension operators are consistent with the general formula\footnote{Here $c_{njk}$ is the coefficient appearing in the CFT 3-point function $\langle {\cal O}_n {\cal O}_j {\cal O}_k \rangle$.  Note that since ${\cal O}_n$ is double-trace,
one could use large-$N$ factorization to compute $c_{njk}$ in terms of two-point functions.  This means $c_{njk}$ is ${\cal O}(1)$
at large $N$, unlike the coefficient of a single-trace three-point function which is ${\cal O}(1/N)$.  Explicit expressions for $c_{njk}$ have indeed been determined \cite{Heemskerk:2009pn,Fitzpatrick:2011dm}, but rather than use these results we concentrate on determining the
combination $a_n^{CFT} c_{njk}$ which is easier in practice and adequate for our purposes.}
\begin{eqnarray}
a^{CFT}_{n}c_{njk}&=&\frac{\lambda}{\pi^{d}}\frac{(-1)^{n}}{\Gamma(\Delta_{j}-\frac{d}{2})\Gamma(\Delta_{k}-\frac{d}{2})}
\frac{\Gamma(\Delta_{j}+n)\Gamma(\Delta_{k}+n)\Gamma(n+\Delta_{j}+\Delta_{k}-\frac{d}{2})}{\Gamma(n+1)\Gamma(2n+\Delta_{j}+\Delta_{k}-\frac{d}{2})}\nonumber\\
&& \times \frac{1}{\Delta_{n}(\Delta_{n}-d)-\Delta_{i}(\Delta_{i}-d)}
\label{coefbulk}
\end{eqnarray}
These coefficients construct a local scalar field (\ref{basicbulkcft}) in the bulk.  By re-summing the tower of higher-dimension operators, in all cases we find that this field obeys the equation of motion
\be
(\nabla^2-\Delta(\Delta-d))\phi_i(x,z) = \frac{\lambda}{N} \phi_j \phi_k(x,z)
\ee
where $\lambda$ is determined by the 3-point coupling of the CFT.  For completeness, in appendix \ref{appendix:bulk} we reproduce the coefficients (\ref{coefbulk}) by solving the radial bulk evolution equations perturbatively.

\noindent{\em Scalar fields coupled to gauge fields}\\
\noindent
We study this in section \ref{sect:gauge}, where we consider a charged scalar field corresponding to an operator of dimension $\Delta$.  We show that the
condition for locality of this bulk scalar with a conserved current on the boundary can be reduced to a scalar system with operators of dimension $\Delta$, $\Delta + 1$, $d - 1$.
This lets us determine the necessary coefficients $a_n^{CFT}$ and show that the bulk scalar obeys the expected equation of motion in holographic gauge.
\be
(\nabla^2-\Delta(\Delta-d))\phi(x,z) = -2i\frac{q}{N}g^{\mu\nu}A_\mu \partial_\nu \phi(x,z)
\ee
Here to ${\cal O}(1/N)$ $g^{\mu\nu}$ is the inverse AdS metric and $q$ is the charge of the scalar field measured in units of $1/N$.

\noindent{\em Scalar fields coupled to gravity}\\
\noindent
In this case, as we show in section \ref{sect:gravity}, the locality condition for a bulk scalar field can be reduced to that of a scalar system with operators of dimension
$\Delta$, $\Delta + 2$, $d$, for which  we determine the coefficients in appendix \ref{appendix:scalargravity}. We show that the resulting bulk scalar obeys
\begin{equation}
(\nabla^2-\Delta(\Delta-d))\phi(x,z)=\kappa g^{\mu\rho} g^{\nu\sigma} h_{\mu \nu} \partial_\rho \partial_\sigma \phi(x,z)
\end{equation}
which is the expected equation of motion to order $\frac{1}{N}$ in holographic gauge.  Here $g^{\mu\nu}$ is the inverse AdS metric, $h_{\mu\nu}$ is the metric perturbation and $\kappa \sim 1/N$ is the gravitational coupling.

An outline of this paper is as follows.  In section \ref{sect:procedure} we study the unwanted non-analyticity in $I_{ijk}$ in more detail and set up the procedure
for restoring bulk locality. In section \ref{sect:d=1} we consider scalar fields in AdS${}_2$ and determine the coefficients of the higher-dimension operators.  In section
\ref{sect:bulkeom} we show that these coefficients imply the expected bulk equations of motion, and in section \ref{sect:redef} we exhibit the freedom in the construction
which corresponds to bulk field redefinitions.  In sections \ref{sect:d=2} and \ref{sect:d=3} we study scalar fields in AdS${}_3$ and $AdS_4$, respectively, and in sections
\ref{sect:gauge} and \ref{sect:gravity} we consider scalar fields coupled to gauge fields and gravity.  Some supporting calculations are gathered in the appendices.

\section{A CFT procedure to restore locality\label{sect:procedure}}
In this section we study the naive bulk-boundary correlator (\ref{NaiveSmear}) in more detail and set up the procedure for restoring locality proposed in \cite{Kabat:2011rz}.

For primary scalars of dimensions $\Delta_i$, $\Delta_j$, $\Delta_k$ the three-point function is determined by conformal invariance up to an overall coefficient.
\be
\label{CFT3ptFunction}
\left\langle {\cal O}_i(x_i) {\cal O}_j(x_j){\cal O}_k(x_k)\right\rangle =\frac{c_{ijk}}{|x_i-x_j|^{\Delta_i+\Delta_j-\Delta_k}|x_i-x_k|^{\Delta_i+\Delta_k-\Delta_j}|x_j-x_k|^{\Delta_j+\Delta_k-\Delta_i}}
\ee
To compare results from the CFT to results from the bulk we need the coefficient $c_{ijk}$. We can get this from known results in AdS/CFT.
To avoid notational confusion we denote the coefficient of the three-point function of single trace operators at leading large $N$ by $\gamma_{ijk}$. Results from AdS/CFT for a cubic coupling $\lambda/N$ in the bulk give\footnote{The case of a different bulk coupling will be addressed section \ref{sect:redef}.} \cite{Freedman:1998tz}
\begin{equation}
\gamma_{ijk}=-\frac{\lambda}{N} \frac{\Gamma[\frac{1}{2}(\Delta_{i}+\Delta_{j}-\Delta_{k})]\Gamma[\frac{1}{2}(\Delta_{i}-\Delta_{j}+\Delta_{k})]\Gamma[\frac{1}{2}(\Delta_{k}+\Delta_{j}-\Delta_{i})]}{2\pi^{d}\Gamma(\Delta_{i}-\frac{d}{2})\Gamma(\Delta_{j}-\frac{d}{2})\Gamma(\Delta_{k}-\frac{d}{2})}\Gamma[\frac{1}{2}(\Delta_{i}+\Delta_{j}+\Delta_{k}-d)]
\label{gijk}
\end{equation}
This result was obtained using Witten's bulk-boundary propagator, and as a result this coefficient does not match what one gets from the bulk by sending
$z \rightarrow 0$ in $\langle \phi_i(x,z) {\cal O}_j {\cal O}_k\rangle$ and looking at the $z^{\Delta_{i}}$ behavior.  Instead, as was noted in \cite{Freedman:1998tz, Giddings:1999qu} and explained in
\cite{Klebanov:1999tb, Harlow:2011ke}, what we get in this limit is
\begin{equation}
\tilde{\gamma}^{(d)}_{ijk}=\frac{\gamma_{ijk}}{2\Delta_{i}-d}\,.
\label{corgijk}
\end{equation}
The factor $2 \Delta_i - d$, which could be absorbed into a wave-function renormalization, accounts for the difference in normalization of the two prescriptions.

The integral of a smearing function against the CFT correlator (\ref{CFT3ptFunction}) was computed in \cite{Kabat:2011rz,Kabat:2012av}, where it was found that
\be
\label{3pointbulk}
\int d^dx' K_i(x,z \vert x') \langle {\cal O}_i(x') {\cal O}_j(y_1) {\cal O}_k(y_2) \rangle = \frac{1}{(y_1 - y_2)^{2\Delta_j}}
\left[\frac{z}{z^2+(x-y_2)^2}\right]^{\Delta_k-\Delta_j} I_{ijk}(\chi)
\ee
Here $I_{ijk}$ is a function of the AdS-invariant distance (\ref{chidef}), explicitly given by
\be
\label{Idef}
I_{ijk}(\chi) = c_{ijk} \left(\frac{1}{\chi-1}\right)^\DS F\big(\,\DS,\,\DS - \frac{d}{2} + 1,\, \Delta_i - \frac{d}{2} + 1,\,\frac{1}{1-\chi}\,\big)
\ee
Here $\Delta_{\textstyle *}=\frac{1}{2}(\Delta_{i}+\Delta_{j}-\Delta_{k})$.\footnote{This combination was denoted $\Delta_0$ in \cite{Kabat:2012av}.}
One can use the transformation
\be
F(\alpha,\beta,\gamma,z)=(1-z)^{-\alpha}F(\alpha,\gamma-\beta,\gamma,\frac{z}{z-1})
\ee
to rewrite $I_{ijk}$ in the equivalent form
\be
\label{Idef2}
I_{ijk}(\chi) = c_{ijk} \left(\frac{1}{\chi}\right)^\DS F\big(\,\DS,\,\Delta_i - \DS,\,\Delta_i - \frac{d}{2} + 1,\,\frac{1}{\chi}\,\big)
\ee
Although the notation makes it somewhat obscure, at spacelike separation the correlator (\ref{3pointbulk}) is invariant under the exchange of
${\cal O}_j(y_1)$ with ${\cal O}_k(y_2)$.

In the following discussion the factor
\be
\label{factor}
\frac{1}{(y_1 - y_2)^{2\Delta_j}} \left[\frac{z}{z^2+(x-y_2)^2}\right]^{\Delta_k-\Delta_j}
\ee
which is present in the correlator (\ref{3pointbulk}) will not play much of a role: it depends on the location of the second (spectator) boundary operator, but it
does not care about the value of $\Delta_i$.  We will restore this factor later, but for now we concentrate on the function $I_{ijk}(\chi)$.
In general $I_{ijk}$ has singularities at $\chi = 0$ and $\chi = 1$ and a branch cut for $0<\chi<1$.  A few examples, appropriate for massless fields in the bulk, are shown in Table \ref{table:I}.

\begin{table}
\begin{center}
\begin{tabular}{l|l}
$d$ \quad & \qquad\quad $I_{ijk}$ \\
\hline \\[-2pt]
1 & \quad $\displaystyle c_{ijk} \left({i \over 2} \log {1 + i \sqrt{\chi - 1} \over 1 - i \sqrt{\chi - 1}} + {\pi \over 2}\right)$ \\[10pt]
2 & \quad $\displaystyle c_{ijk} \log {\chi \over \chi - 1}$ \\[10pt]
3 & \quad $\displaystyle 3 c_{ijk} \left({1 \over \sqrt{\chi - 1}} - {i \over 2} \log {1 + i \sqrt{\chi - 1} \over 1 - i \sqrt{\chi - 1}} - {\pi \over 2}\right)$ \\[15pt]
4 & \quad $\displaystyle 2 c_{ijk} \left({1 \over \chi - 1} - \log {\chi \over \chi - 1}\right)$
\end{tabular}
\end{center}
\caption{$I_{ijk}$ in some low-dimensional examples of AdS${}_{d+1}$ / CFT${}_d$.  The examples are for massless fields in the bulk, that is,
for $\Delta_i = \Delta_j = \Delta_k = d$.\label{table:I}}
\end{table}

The behavior near $\chi = 1$ was studied in section 4 of \cite{Kabat:2012av}.  If $d$ is an odd integer (that is, in even AdS) it was found that generically
$I_{ijk}$ has an expansion about $\chi = 1$ of the form
\begin{equation}
\label{oddd}
I_{ijk} = (\chi-1)^{1 - \frac{d}{2}} \sum_{k=0}^\infty \alpha_k (\chi-1)^k
\end{equation}
Due to the square root branch cut, this gives a non-zero commutator at spacelike separation.  If $d$ is an even integer the expansion generically has the form
\begin{equation}
\label{evend}
\sum_{k=0}^{d/2} \beta_k (\chi-1)^{k + 1 - \frac{d}{2}} + \ln(\chi-1) \sum_{k=0}^{\infty} \gamma_k (\chi - 1)^{k}
\end{equation}
Again this results in a non-zero commutator at spacelike separation.  There are exceptions to these general rules, but they won't matter for our purposes.\footnote{In
particular if $\DS$ or $\Delta_i - \DS$ is a non-positive integer the hypergeometric series in (\ref{Idef2}) terminates and $I_{ijk}$ is analytic about $\chi = 1$.  But in these cases (\ref{gijk}) diverges and the
bulk interaction vertex is subtle.  For further discussion see \cite{Liu:1999kg}.}
Thus generically we find that (\ref{3pointbulk}) gives a non-zero commutator at spacelike separation.

It is important to note that, as shown in (\ref{oddd}) and (\ref{evend}), the form of the expansion about $\chi = 1$ only cares about the spacetime dimension
$d$ of the CFT.  In particular all information about the operator dimension $\Delta_i$ is hidden inside the expansion coefficients $\alpha_k$, $\beta_k$, $\gamma_k$.  This means that
if we have an infinite tower of operators ${\cal O}_n$ at our disposal, all of which have non-trivial 3-point functions with ${\cal O}_j$ and ${\cal O}_k$, to ${\cal O}(1/N)$ we can
try to cancel the unwanted non-analyticity by constructing a bulk field of the form
\begin{equation}
\phi_{i}(x,z)=\int d^dx' K_{\Delta_i}(x,z \vert x') {\cal O}_i(x') + {1 \over N} \sum_n a^{CFT}_{n} \int d^dx' K_{\Delta_n}(x,z \vert x') {\cal O}_{n}(x')
\label{basicbulkcft2}
\end{equation}
We fix the coefficients $a^{CFT}_{n}$ to cancel any unwanted non-analyticity at $\chi > 0$.  In all dimensions there is a branch cut, but as can be seen in (\ref{evend}) in even $d \geq 4$ there are
also poles at $\chi = 1$ that must be canceled.  We study this in appendix \ref{appendix:polecut} and show that cancellation of the cut automatically implies
cancellation of the poles.  So for now we concentrate on the cut and study the discontinuity in $I_{ijk}$ in more detail.  The
discontinuity of a hypergeometric function across its cut is
 \begin{eqnarray}
&& F(\alpha,\beta,\gamma,z+i\epsilon) - F(\alpha,\beta,\gamma,z-i\epsilon) \\
\nonumber
&& = \frac{2\pi i \Gamma(\gamma) (z-1)^{\gamma-\alpha-\beta}} {\Gamma(\alpha) \Gamma(\beta) \Gamma(1+\gamma-\alpha-\beta)}
F(\gamma-\alpha,\gamma-\beta,1+\gamma-\alpha-\beta,1-z)
\end{eqnarray}
This allows us to determine the discontinuity across the unwanted cut of $I_{ijk}$.  Dropping a factor $2 \pi i$ it is given by
\begin{equation}
I_{discont}^{ijk}=\frac{c_{ijk}}{(1-\chi)^{\frac{d}{2}-1}} \frac{\Gamma(\Delta_i-\frac{d}{2}+1)}{\Gamma(\DS)\Gamma(\Delta_i-\DS)\Gamma(2-\frac{d}{2})} F\big(\,\DS-\frac{d}{2}+1,\,1+\DS-\Delta_i,\,2-\frac{d}{2},\,1-\chi\,\big)
\label{discont}
\end{equation}
For even $d \geq 4$ one must further use
\begin{equation}
\lim_{\gamma \rightarrow -n} \frac{F(\alpha, \beta, \gamma, z)}{\Gamma(\gamma)} =
\frac{(\alpha)_{n+1} (\beta)_{n+1}}{(n+1)!}z^{n+1}F(\alpha+n+1,\, \beta+n+1,\,n+2,\,z)
\label{Evendgeq4Discont}
\end{equation}
The basic condition to cancel the unwanted branch cut is then\footnote{We dropped the $jk$ indices to avoid later clutter.}
\be
\label{basiccondition}
I_{discont}^{\Delta_i} = - {1 \over N} \sum_n a^{CFT}_n \, I_{discont}^{\Delta_n}
\ee
In the rest of this paper we solve this equation in numerous examples and use the results to derive the bulk equations of motion.

\section{CFT computation in $d=1$\label{sect:d=1}}
We start with the simplest example of a massless scalar field in AdS${}_2$, dual to a primary scalar field which has a non-trivial 3-point function with itself and whose dimension $\Delta=1$.
Equation (\ref{discont}) gives the discontinuity that we wish to cancel as
\begin{equation}
I^{1}_{discont}=\tilde{\gamma}^{(d=1)}_{111}\frac{(1-\chi)^{\frac{1}{2}}}{\Gamma(\frac{3}{2})}\frac{\Gamma(\frac{3}{2})}{\Gamma^{2}(\frac{1}{2})}F(1,\frac{1}{2},\frac{3}{2},1-\chi)
\end{equation}
For this case equation (\ref{corgijk}) gives $\tilde{\gamma}^{(d=1)}_{111}=-\frac{\lambda}{2\pi N}$.

It's convenient to work in terms of Legendre functions $Q_\nu(x)$, $P_\nu(x)$.  These can be represented by
\begin{eqnarray}
Q_{\nu}(x)&=&-\frac{\sqrt{\pi}\sin(\frac{\nu \pi}{2})\Gamma(\frac{\nu +1}{2})}{2\Gamma(\frac{\nu}{2} +1)}F(\frac{\nu+1}{2},\frac{-\nu}{2},\frac{1}{2},x^2) \nonumber\\
&&+\frac{\sqrt{\pi}\cos(\frac{\nu \pi}{2})\Gamma(\frac{\nu}{2}+1)}{\Gamma(\frac{\nu +1}{2})}xF(\frac{\nu}{2}+1,\frac{1-\nu}{2},\frac{3}{2},x^2)\nonumber\\
P_{\nu}(x)&=&\frac{\cos(\frac{\nu \pi}{2})\Gamma(\frac{\nu +1}{2})}{\sqrt{\pi}\Gamma(\frac{\nu}{2} +1)}F(\frac{\nu+1}{2},\frac{-\nu}{2},\frac{1}{2},x^2) \nonumber \\
&&+\frac{2\sin(\frac{\nu \pi}{2})\Gamma(\frac{\nu}{2}+1)}{\sqrt{\pi}\Gamma(\frac{\nu +1}{2})}xF(\frac{\nu}{2}+1,\frac{1-\nu}{2},\frac{3}{2},x^2)
\label{legendre1}
\end{eqnarray}
Here the Legendre function $Q_\nu(x)$ is defined to have branch cuts
along $(-\infty,-1) \cup (1,\infty)$, for example $Q_0(x) = {1 \over 2} \log {1 + x \over 1 - x}$.
From these expressions we identify
\begin{equation}
I^{1}_{discont}=-\frac{\lambda}{2\pi^2 N}Q_{0}(\sqrt{1-\chi})
\end{equation}
On the other hand the double-trace operators we can make from two operators of dimension one and $2n$ derivatives have dimensions $\Delta_n = 2n + 2$,
which give according to (\ref{discont}) a discontinuity of the form
\begin{equation}
I^{2+2n}_{discont}=c_{n11}\frac{(1-\chi)^{1/2}}{\Gamma(3/2)}\frac{\Gamma(2n+2\frac{1}{2})}{\Gamma^{2}(n+1)}F(n+\frac{3}{2},-n,\frac{3}{2},1-\chi)
\end{equation}
or equivalently
\begin{equation}
I^{2+2n}_{discont}=(-1)^{n} c_{n11}\frac{\Gamma(2n+\frac{5}{2})}{\Gamma(n+\frac{3}{2})\Gamma(n+1)}P_{2n+1}(\sqrt{1-\chi})
\label{dobtrc1}
\end{equation}
So we need to find a way to write $Q_{0}(x)$ as a sum of Legendre polynomials with odd index.
Using the integrals
\begin{eqnarray}
&&\int_{-1}^{1}P_{m}(x)P_{n}(x)=\delta_{nm}\frac{2}{2n+1}\nonumber\\
&&\int_{-1}^{1}P_{m}(x)Q_{\nu}(x)=\frac{1-\cos\pi(\nu-m)}{(m-\nu)(m+\nu+1)}\label{legendre2}\\
&&\int_{-1}^{1}P_{m}(x)P_{\nu}(x)=-\frac{2}{\pi}\frac{\sin\pi(\nu-m)}{(m-\nu)(m+\nu+1)}\nonumber
\end{eqnarray}
we see that we can write
\begin{equation}
Q_{0}(x)=\sum_{k=0}^{\infty}\frac{4k+3}{(2k+1)(2k+2)}P_{2k+1}(x)
\label{q0ps}
\end{equation}

This means the unwanted non-analyticity in the correlator can indeed be canceled by redefining the bulk field as in (\ref{basicbulkcft2}), namely
\begin{equation}
\phi(x,z) = \int dx' K_{1}(x,z \vert x') {\cal O}(x') + \frac{1}{N} \sum_{n} a^{CFT}_{n} \int dx' K_{\Delta_n}(x,z \vert x') {\cal O}_{n}(x')
\label{cftoper}
\end{equation}
The coefficients $a_n^{CFT}$ should be chosen to satisfy the cancellation condition (\ref{basiccondition}), which given (\ref{q0ps}) means
that we can set
\begin{equation}
a^{CFT}_{n}c_{n11}=\frac{\lambda (-1)^{n}\Gamma(n+\frac{3}{2})\Gamma(n+1)}{\pi^2(2n+1)(2n+2)\Gamma(2n+\frac{3}{2})}
\end{equation}
Since in this case $\Delta_i=\Delta_j=\Delta_k=1$ and $\Delta_n=2n+2$, this agrees with the general form (\ref{coefbulk}).

\subsection{Generalization}
We now generalize to the case of a primary operator of dimension $\Delta_{\nu}$ (which is not necessarily an integer) and two primary operators of dimension one (so $\Delta_{j}=\Delta_k =1$). In AdS${}_2$ it is convenient to set $\Delta_\nu = \nu + 1$, since $\nu$ will become the order of a Legendre function.

We wish to uplift the operator of dimension $\Delta_{\nu}$ to a bulk field.  From (\ref{discont}) the discontinuity in the 3-point function of operators of dimensions $\Delta_{\nu}$ and $\Delta_{j}=\Delta_{k}=1$ is
\begin{equation}
I_{discont}^{\Delta_{\nu}}=\tilde{\gamma}^{(d=1)}_{\Delta_{\nu}11}(1-\chi)^{1/2}\frac{2\Gamma(\nu+\frac{3}{2})}{\sqrt{\pi}\Gamma^{2}(\frac{\nu+1}{2})}F(\frac{\nu+2}{2},\frac{1-\nu}{2},\frac{3}{2},1-\chi)
\label{d1nu}
\end{equation}
with $\tilde{\gamma}^{(d=1)}_{\Delta_{\nu}11}$ given by (\ref{corgijk}) to be
\begin{equation}
\tilde{\gamma}^{(d=1)}_{\Delta_{\nu}11}=-\frac{\lambda}{4\pi N}\frac{\Gamma^2(\frac{\nu +1}{2})\Gamma(\frac{1-\nu}{2})}{\Gamma(\nu+\frac{3}{2})\Gamma^2(\frac{1}{2})}\Gamma(\frac{\nu}{2}+1)=-\frac{\lambda}{4\pi N}\frac{\Gamma(\frac{\nu +1}{2})\Gamma(\frac{\nu}{2}+1)}{\Gamma(\nu+\frac{3}{2})\cos \frac{\pi \nu}{2}}
\end{equation}
Equation (\ref{d1nu}) can be written using (\ref{legendre1}) as
\begin{equation}
\tilde{\gamma}^{(d=1)}_{\Delta_{\nu}11}\frac{2\Gamma(\nu+\frac{3}{2})}{\pi \Gamma(\frac{\nu+1}{2})\Gamma(\frac{\nu}{2}+1)}f_{\nu}(\sqrt{1-\chi})
\label{fnueqn}
\end{equation}
where
\begin{equation}
f_{\nu}(x)=\frac{\pi}{2}\sin(\frac{\pi \nu}{2})P_{\nu}(x)+\cos(\frac{\pi \nu}{2})Q_{\nu}(x)
\end{equation}

The higher-dimension double-trace operators we can add are the same as before, so they give rise to the same discontinuity (\ref{dobtrc1}). 
So we need to find a way of writing $f_{\nu}$ as a sum of Legendre polynomial of odd order.
Using (\ref{legendre2}) we have
\begin{equation}
\int_{-1}^{1}P_{m}(x)f_{\nu}(x)=
\left\lbrace
\begin{array}{cl}
\frac{2\cos(\frac{\pi \nu}{2})}{(m-\nu)(m+\mu+1)} & \hbox{\rm $m$ odd} \\[5pt]
0 & \hbox{\rm $m$ even}
\end{array}
\right.
\end{equation}
so we can write
\begin{equation}
f_{\nu}(x)=\cos(\frac{\pi \nu}{2})\sum_{k=0}^\infty\frac{4k+3}{(2k+1-\nu)(2k+2+\nu)}P_{2k+1}(x)\,.
\label{fnu}
\end{equation}
To cancel the commutator at bulk spacelike separation we need to satisfy (\ref{basiccondition}), which means
\begin{equation}
a^{CFT}_{n}c_{n}=\frac{\lambda}{\pi^2} \frac{\Gamma(n+\frac{3}{2})\Gamma(n+1)}{\Gamma(2n+\frac{3}{2})}\frac{(-1)^n}{(2n+\nu+1)(2n+2+\nu)}
\label{NuOneOne}
\end{equation}
Note that since $\Delta_{n}=2+2n$ we have
\begin{equation}
\frac{1}{(2n+\nu+1)(2n+2+\nu)}=\frac{1}{\Delta_{n}(\Delta_{n}-1)-\Delta_{\nu}(\Delta_{\nu}-1)}
\end{equation}
This means (\ref{NuOneOne}) again agrees with the general result (\ref{coefbulk}).

\subsection{Further generalization}
We now treat the case where we have an operator of dimension $\Delta_{\nu} = \nu + 1$ which has a non-trivial 3-point function with two operators of dimension $\Delta_{j}=\Delta_{k}=l \in {\mathbb N}$. We want to promote the operator of dimension $\Delta_{\nu}$ to a local bulk field.

Note that the value of $\DS = \Delta_\nu / 2$ does not depend on $l$, so the discontinuity in the initial 3-point function is simply given by (\ref{fnueqn})
with the replacement $\tilde{\gamma}^{(d=1)}_{\Delta_{\nu}11} \rightarrow \tilde{\gamma}^{(d=1)}_{\Delta_{\nu}ll}$.
However in this case the double-trace operators we can build have dimension $\Delta_{n}=2l+2n$ and will produce discontinuities involving
the Legendre polynomials
$P_{2n+2l-1}(x)$. This means that in attempting to write a formula like (\ref{fnu}) we do not have all the required Legendre polynomials. We can
nevertheless proceed as follows. The Legendre functions $P_{\nu}(x)$ and $Q_{\nu}(x)$ both obey the differential equation
\begin{equation}
L_{\mu}g_{\nu}=\left[-\frac{d}{dx}(1-x^2)\frac{d}{dx}-\mu(\mu+1)\right]g_{\nu}=[\nu(\nu+1)-\mu(\mu+1)]g_{\nu}
\label{operator}
\end{equation}
Thus we can act with $\prod_{i=1}^{l-1}L_{2i-1}$
on both sides of (\ref{fnu}) to eliminate terms on the right hand side involving $P_{m}$ for $m<2l-1$. On the left hand side we just get a multiplicative factor of 
\begin{equation}
h_{\nu,l}=\prod_{i=1}^{l-1}(\nu(\nu+1)-(2i-1)2i)=(-1)^{l-1}4^{l-1}\frac{\Gamma(\frac{\nu}{2}+l)\Gamma(l-\frac{\nu}{2}+\frac{1}{2})}{\Gamma(\frac{\nu}{2}+1)\Gamma(\frac{1}{2}-\frac{\nu}{2})}
\end{equation}
Thus we get a modified version of (\ref{fnu}) which is suitable for our purposes.
\begin{equation}
f_{\nu}(x)=\frac{1}{h_{\nu,l}}\cos(\frac{\pi \nu}{2})\sum_{k=l-1}^{\infty}\frac{(4k+3)\prod_{i=1}^{l-1}[(2k+1)(2k+2)-2i(2i-1)]}{(2k+1-\nu)(2k+2+\nu)}P_{2k+1}(x)
\label{fnul}
\end{equation}
This should be regarded as a formal expression rather than as a convergent sum (we will encounter several such formal expressions as we go along).  But
this formal expression will enable us to compute the $a_n^{CFT}$ coefficients which are necessary to
describe local fields in the bulk.  To simplify this expression note that the product can be evaluated by
\begin{eqnarray}
&  & \prod_{i=1}^{l-1}\big[(2k+1)(2k+2)-2i(2i-1)\big] \nonumber \\
&=& \prod_{i=1}^{l-1}(2k-2i+2)(2k+2i+1) \nonumber \\
&=& 4^{l-1}\frac{\Gamma(n+l)\Gamma(n+2l-\frac{1}{2})}{\Gamma(n+1)\Gamma(n+l+\frac{1}{2})}
\label{factor1}
\end{eqnarray}
where in the last line we set $k=n+l-1$.  It's also helpful to note that
\begin{equation}
(2k+1-\nu)(2k+2+\nu) = \Delta_{n}(\Delta_{n}-1)-\Delta_{\nu}(\Delta_{\nu}-1)
\end{equation}

Now the discontinuity coming from a double-trace operator of dimension $\Delta_{n}=2n+2l$ is
\begin{equation}
I_{discont}^{2n+2l}=c_{nll}(-1)^{n+l+1}\frac{\Gamma(2n+2l+\frac{1}{2})}{\Gamma(n+l)\Gamma(n+l+\frac{1}{2})}P_{2n+2l-1}(\sqrt{1-\chi})
\end{equation}
So the condition for the discontinuities to cancel (\ref{basiccondition}) gives
\begin{eqnarray}
\label{gencofd1}
& &a_{n}^{CFT}c_{nll}=-N\tilde{\gamma}^{(d=1)}_{\Delta_{\nu} ll}\frac{4}{\pi}\frac{\cos(\frac{\pi \nu}{2})\Gamma(\nu+\frac{3}{2})\Gamma(\frac{1}{2}-\frac{\nu}{2})}{\Gamma(\frac{\nu}{2}+l)\Gamma(\frac{\nu+1}{2})\Gamma(l-\frac{\nu}{2}-\frac{1}{2})}\times\\
& & \frac{(-1)^{n}\Gamma^2(n+l)\Gamma(n+2l-\frac{1}{2})}{\Gamma(n+1)\Gamma(2n+2l-\frac{1}{2})}
\frac{1}{(2n+2l)(2n+2l-1)-(\Delta_{\nu}-1)\Delta_{\nu}}
\nonumber
\end{eqnarray}
From (\ref{corgijk}) we have
\begin{equation}
\tilde{\gamma}^{(d=1)}_{\Delta_{\nu} ll}=-\frac{\lambda}{4\pi N}\frac{\Gamma^{2}(\frac{\nu+1}{2})\Gamma(l-\frac{\nu}{2}-\frac{1}{2})\Gamma(\frac{\nu}{2}+l)}{\Gamma(\nu +\frac{3}{2})\Gamma^2(l-\frac{1}{2})}
\end{equation}
So putting it all together, and using $\Gamma(\frac{1}{2}+x)\Gamma(\frac{1}{2}-x)=\frac{\pi}{\cos \pi x}$, this again agrees with the general result
(\ref{coefbulk}).
 
\subsection{Odd $\nu$}
The formulas in the previous subsection are valid as long as $\nu$ is not odd, so we treat this case separately. This case is also interesting since it
exhibits some aspects that were a bit hidden before.
We illustrate this for $\Delta_{\nu}=\nu+1=2$ and $\Delta_{j}=\Delta_{k}=l=2$.

We start with a three-point function of operators of dimension two, so the discontinuity we want to cancel is
\begin{equation}
\tilde{\gamma}^{(d=1)}_{222}\frac{3}{2}P_{1}(\sqrt{1-\chi})
\end{equation}
where
\begin{equation}
\tilde{\gamma}^{(d=1)}_{222}=-\frac{\lambda}{4\pi N \Gamma^{2}(\frac{3}{2})}\,.
\end{equation}
The higher-dimension operators we have at our disposal are of dimension $\Delta_{n}=4+2n$ and give rise to a discontinuity
\begin{equation}
c_{n22}(-1)^{n+1}\frac{\Gamma(2n+\frac{9}{2})}{\Gamma(n+2)\Gamma(n+\frac{5}{2})}P_{2n+3}(\sqrt{1-\chi})
\end{equation}
So we need to write $P_{1}(x)$ as an infinite sum of $P_{2n+3}(x)$. This does not seem possible, but let us proceed formally. We start with (\ref{q0ps})
and act with $L_{0}$ on both sides to obtain
\begin{equation}
\sum_{k=0}^\infty(4k+3)P_{2k+1}(x)=0
\end{equation}
(another formal expression!).  From this we can write
\begin{equation}
\tilde{\gamma}^{(d=1)}_{222}\frac{3}{2}P_{1}(\sqrt{1-\chi})=-\frac{1}{2}\tilde{\gamma}^{(d=1)}_{222}\sum_{k=1}^\infty(4k+3)P_{2k+1}(x)
\end{equation}
Using this to cancel the discontinuity as in (\ref{basiccondition}), we find (setting $k=n+1$)
\begin{equation}
a^{CFT}_{n}c_{n22}=\lambda (-1)^{n}\frac{\Gamma(n+2)\Gamma(n+2\frac{1}{2})}{4\pi \Gamma^{2}(\frac{3}{2})\Gamma(2n+3\frac{1}{2})}
\end{equation}
Using  $(2n+4)(2n+3)-2=4(n+1)(n+\frac{5}{2})$ we can write this as 
\begin{equation}
a^{CFT}_{n}c_{n22}=\lambda (-1)^{n}\frac{\Gamma^{2}(n+2)\Gamma(n+3\frac{1}{2})}{\pi \Gamma^{2}(\frac{3}{2})\Gamma(n+1)\Gamma(2n+3\frac{1}{2})}\frac{1}{(2n+4)(2n+3)-2}
\end{equation}
which again matches the general result (\ref{coefbulk}).

\section{Recovering bulk equations of motion\label{sect:bulkeom}}
In the previous section we showed that to ${\cal O}(1/N)$ we could build a local field in the bulk by adding an infinite tower of smeared
double-trace operators to the CFT definition a bulk field.  We chose the coefficients of these double-trace operators so that, by construction,
we obtained a field which respects bulk locality.  However the resulting representation (\ref{basicbulkcft}) is not how one usually thinks of a field in AdS, so we'd like
to understand
how our construction is related to the standard ideas of bulk effective field theory.  Here we address this by showing that the bulk
field we have built up obeys an equation of motion, which in fact is nothing but the usual Heisenberg equation of motion in the bulk.
This means the requirement of bulk locality is sufficient to derive the bulk equations of motion from CFT correlators, not only in principle, but in practice.

First remember that a local bulk field can be expressed in terms of the CFT as
\begin{equation}
\phi(x,z) = \int d^dx' K_{\Delta}(x,z \vert x') {\cal O}(x') + \frac{1}{N} \sum_{n} a^{CFT}_{n} \int d^dx' K_{\Delta_n}(x,z \vert x') {\cal O}_{n}(x')
\end{equation}
To read off the bulk equations of motion we act with $\nabla^2-\Delta(\Delta-d)$ to get
\begin{equation}
\big(\nabla^2-\Delta(\Delta-d)\big)\phi(x,z)=\frac{1}{N}\sum_{n} a^{CFT}_{n}\big[(\Delta_{n}(\Delta_{n}-d)-\Delta(\Delta-d)\big] \int d^dx' K_{\Delta_n}(x,z \vert x') {\cal O}_{n}(x')
\label{eomcft}
\end{equation}
This is fine, of course, but on the right hand side of this expression we have an infinite sum of smeared CFT operators.  How do we identify what this means in bulk terms?
The approach we will take is to insert the right hand side of (\ref{eomcft}) in a CFT correlator and use the result to figure out the bulk interpretation.
In fact we will put the right hand side of (\ref{eomcft}) into a correlator with ${\cal O}_{j}(y_1)$ and ${\cal O}_{k}(y_2)$.  This approach is particularly useful because the correlator
only depends on the combination $a_{n}^{CFT}c_{njk}$, which is exactly the combination we computed in the previous section.

As we will see shortly the following result is very useful. Since we only need to study the operator on the right hand side of (\ref{eomcft}) at leading large $N$,
we can use large-$N$ factorization to obtain\footnote{The superscript
${}^{(0)}$ means that $\phi^{(0)}$ is a free field, obtained by zeroth order smearing of the CFT operator.  The same notation appears in appendix \ref{appendix:bulk}.}
\begin{eqnarray}
&&\langle\phi^{(0)}_{j}\phi^{(0)}_{k}(x,z){\cal O}_{j}(y_1){\cal O}_{k}(y_2)\rangle =\langle\phi^{(0)}_{j}(x,z){\cal O}_{j}(y_1)\rangle \langle\phi^{(0)}_{k}(x,z){\cal O}_{k}(y_2)\rangle \nonumber\\
&&= d_{jk}\left(\frac{z}{(x-y_1)^2+z^2}\right)^{\Delta_j}\left(\frac{z}{(x-y_2)^2+z^2}\right)^{\Delta_k}
\label{Maincoef2}
\end{eqnarray}
where
\begin{equation}
d_{jk}=\frac{1}{\pi^{d}}\frac{\Gamma(\Delta_{j})\Gamma(\Delta_{k})}{\Gamma(\Delta_{j}-\frac{d}{2})\Gamma(\Delta_{k}-\frac{d}{2})}
\label{Maindjk}
\end{equation}

Recall that the 3-point function of any of the smeared ${\cal O}_{n}$ can be obtained from (\ref{3pointbulk}), (\ref{Idef}). Assuming $d$ is odd\footnote{For an even dimensional example see section \ref{EvendEOM}.} we find it useful to rewrite the result using the identity
\begin{eqnarray}
& & F(a,b,c,z)=(-z)^{-a}\frac{\Gamma(c)\Gamma(b-a)}{\Gamma(c-a)\Gamma(b)}F(a,a-c+1,a-b+1,\frac{1}{z})\nonumber\\
& & \qquad + (-z)^{-b}\frac{\Gamma(c)\Gamma(a-b)}{\Gamma(c-b)\Gamma(a)}F(b,b-c+1,b-a+1,\frac{1}{z})
\label{hypertrans1}
\end{eqnarray}
Applying this to the $\chi$-dependent terms in (\ref{Idef}) one gets
\begin{eqnarray}
\label{anapartd}
& & \frac{\Gamma(\Delta_i-\frac{d}{2}+1)\Gamma(1-\frac{d}{2})}{\Gamma(\Delta_{i}-\DS-\frac{d}{2}+1)\Gamma(\DS-\frac{d}{2}+1)} F(\DS,\frac{d}{2}+\DS-\Delta_i,\frac{d}{2},1-\chi)\\
& +&\frac{1}{(\chi-1)^{\frac{d}{2}-1}} \frac{\Gamma(\Delta_i-\frac{d}{2}+1)\Gamma(\frac{d}{2}-1)}{\Gamma(\DS)\Gamma(\Delta_i-\DS)} F(\DS-\frac{d}{2}+1,1+\DS-\Delta_i,2-\frac{d}{2},1-\chi)\nonumber
\end{eqnarray}
Note that the first term is analytic about $\chi = 1$.  The second term is singular at $\chi = 1$ and in odd $d$ has a branch cut for $\chi \in (0,1)$.
(In fact, up to a numerical factor, the second term is exactly the discontinuity shown in (\ref{discont}).)  But the left hand side of (\ref{eomcft}) is constructed to have
correlators which are analytic at spacelike separation.  So the right hand side must also be analytic, which means the second term in (\ref{anapartd}) will not contribute after
the sum over $n$ is carried out.  Thus to determine the equation of motion, we only need to sum the contribution of the first term in (\ref{anapartd}).

For simplicity we consider the case $\Delta_j = \Delta_k=l$ (a different case is treated in appendix \ref{appendix:scalargauge}). The terms in (\ref{3pointbulk}) that contribute to the sum are
\begin{eqnarray}
\label{contribute}
c_{nll}\frac{1}{(y_1 -y_2)^{2\Delta_{j}}} \frac{\Gamma(\Delta_{n}-\frac{d}{2}+1)\Gamma(1-\frac{d}{2})}{\Gamma^2(\frac{\Delta_{n}-d}{2}+1)}F(\frac{\Delta_{n}}{2},-\frac{\Delta_{n}}{2}+\frac{d}{2},\frac{d}{2},1-\chi)
\end{eqnarray}
where $\Delta_{n}=2n+2l$.
In this section we consider some examples for $d=1$.  Similar examples for $d=2$ and $d=3$ can be found in sections \ref{sect:d=2} and \ref{sect:d=3}. First we use (\ref{legendre1})
to write
\begin{equation}
\label{FQ}
F(\frac{\Delta_{n}}{2},-\frac{\Delta_{n}}{2}+\frac{1}{2},\frac{1}{2},1-\chi)
=\frac{(-1)^{n+\Delta}2\Gamma(\frac{\Delta{n}}{2}+\frac{1}{2})}{\sqrt{\pi}\Gamma(\frac{\Delta_{n}}{2})}Q_{\Delta_{n}-1}(\sqrt{1-\chi})
\end{equation}
Next recall our result (\ref{gencofd1}) for $a_{n}^{CFT}c_{njk}$.  Since $\Delta_j=\Delta_k=l$ and $\Delta_n=2l+2n$ we have
\begin{equation}
\label{ac}
[M_{n}^{2}-\Delta_{\nu}(\Delta_{\nu}-d)]a_{n}^{CFT}c_{nll}=\frac{\lambda}{\pi\Gamma^2(l-\frac{1}{2})}\frac{(-1)^n\Gamma^2(n+l)
\Gamma(n+2l-\frac{1}{2})}{\Gamma(n+1)\Gamma(2n+2l-\frac{1}{2})}
\end{equation}
where $M_n^2 = \Delta_n(\Delta_n - d)$.

Now let's evaluate the right hand side of (\ref{eomcft}).  First we consider the case $l=1$ . This means $\Delta_{n}=2n+2$, so we need to sum
\begin{equation}
-\frac{\lambda}{\pi^2}\frac{1}{(y_1 -y_2)^{2}}\sum_{n=0}^{\infty}(4n+3)Q_{2n+1}(\sqrt{1-\chi})
\end{equation}
From the known sums \cite{Prudnikov:1990:IS}
\begin{equation}
\sum_{k=0}^{\infty}(\pm1)^{k}(2k+1)Q_{k}(x)=\frac{1}{x \mp 1}
\label{basican}
\end{equation}
we can get a sum over odd-order Legendre functions
\begin{equation}
\sum_{n=0}^{\infty}(4n+3)Q_{2n+1}(x)=\frac{1}{x^2-1}
\label{basican1}
\end{equation}
Thus we find that
\begin{eqnarray}
& &-\frac{\lambda}{\pi^2}\frac{1}{(y_1 -y_2)^{2}}\sum_{n}(4n+3)Q_{2n+1}(\sqrt{1-\chi})\nonumber\\
& &=\frac{\lambda}{\pi^2}\left(\frac{z}{(x-y_1)^2+z^2}\right)\left(\frac{z}{(x-y_2)^2+z^2}\right)
\end{eqnarray}
According to (\ref{Maincoef2}) and (\ref{Maindjk}) this is just
\begin{equation}
\lambda \langle\phi_{j}^{(0)}\phi_{k}^{(0)}(x,z) {\cal O}(y_1){\cal O}(y_2)\rangle .
\end{equation}
Thus we can identify
\begin{equation}
\sum_{n} a^{CFT}_{n}[(\Delta_{n}(\Delta_{n}-d)-\Delta(\Delta-d)] \int d^dx' K_{\Delta_n}(x,z \vert x'){\cal O}_{n}(x')=\lambda \phi_{j}^{(0)}\phi_{k}^{(0)}(x,z)
\label{anares}
\end{equation}

Next we consider the case of general integer $l$. According to (\ref{contribute}), (\ref{FQ}), (\ref{ac}) we need to do the sum
\begin{equation}
\sum_{n=0}^{\infty}\frac{\lambda}{(y_1-y_2)^{2l}}\frac{(-1)^l(4n+4l-1)}{\pi \Gamma^{2}(l-\frac{1}{2})}\frac{\Gamma(n+l)\Gamma(n+2l-\frac{1}{2})}{\Gamma(n+1)\Gamma(n+l+\frac{1}{2})}Q_{2n+2l-1}(\sqrt{1-\chi})
\label{gend1sum}
\end{equation}
Note that
\begin{eqnarray}
& &L_{2l-3}\cdots L_1 \sum_{k=0}^{\infty}(4k+3)Q_{2k+1}(x)\nonumber\\
& & = 4^{l-1}\sum_{n=0}^{\infty}\frac{\Gamma(n+l)\Gamma(n+2l-\frac{1}{2})}{\Gamma(n+1)\Gamma(n+l+\frac{1}{2})}(4n+4l-1)Q_{2n+2l-1}(x)\nonumber
\end{eqnarray}
Also using
\begin{equation}
L_{2i-3}\frac{1}{(1-x^2)^{i-1}}=-\frac{4(i-1)^2}{(1-x^2)^{i}}
\end{equation}
we get
\begin{equation}
L_{2l-3}\cdots L_1 \frac{1}{x^2-1}=4^{l-1}\frac{\Gamma^2(l) (-1)^l}{(1-x^2)^l}
\end{equation}
Using this and (\ref{basican1}) we can do the sum (\ref{gend1sum}) and find it to be
\begin{equation}
\frac{\lambda}{(y_1-y_2)^{2l}}\frac{\Gamma^{2}(l)}{\pi\Gamma^{2}(l-\frac{1}{2})}\frac{1}{\chi^{l}}=\frac{\lambda \Gamma^{2}(l)}{\pi\Gamma^{2}(l-\frac{1}{2})}\left(\frac{z}{(x-y_1)^2+z^2}\right)^{l}\left(\frac{z}{(x-y_2)^2+z^2}\right)^{l}
\end{equation}
According to (\ref{Maincoef2}) and (\ref{Maindjk}), we again find that (\ref{anares}) is satisfied.
This means the bulk field we have defined in the CFT, which was constructed to satisfy bulk locality, in fact to ${\cal O}(1/N)$ obeys the equation of motion
\begin{equation}
\label{Phi3eom}
(\nabla^2-m^2)\phi_{i}(x,z)=\frac{\lambda}{N}\phi_{j}^{(0)}\phi_{k}^{(0)}(x,z)
\end{equation}
Thus our bulk field has a local cubic coupling, which is the only interaction we could hope to see at this order in $\frac{1}{N}$.  One might ask why we found a simple $\phi^3$ coupling,
instead of a more complicated coupling involving derivatives.  We address this in the next section.

\section{Field redefinitions\label{sect:redef}}
In the previous section, when we summed the contribution of the higher-dimension operators to obtain the bulk equations of motion, we always found a simple $\phi^3$ interaction in the bulk.
This may strike the reader as somewhat surprising.  Why did we find a $\phi^3$ interaction, instead of some more complicated cubic vertex involving derivatives?  To sharpen the puzzle, note that the CFT
three-point function is determined up to an overall coefficient by conformal invariance.  So aside from an overall coupling constant, to ${\cal O}(1/N)$ the CFT doesn't seem to know what type of cubic
interaction vertex is present in the bulk.  But if the CFT doesn't know about bulk interactions, how could we hope to recover bulk equations of motion from CFT correlators?

To resolve these questions it's important to note that certain ambiguities are present in the construction on both the CFT and bulk sides.
\begin{itemize}
\item
On the CFT side we uplifted CFT operators to mimic local bulk fields by requiring microcausality, that is, by requiring that correlators be analytic at spacelike separation.  But this uplift procedure
is ambiguous, because we are always free to add something to the CFT definition of a bulk observable which doesn't spoil analyticity at spacelike separation.
\item
From the bulk perspective, the definition of a bulk field is ambiguous because we are always free to make local field redefinitions in the bulk.
\end{itemize}
It's easy to see that these two ambiguities are equivalent to each other.
\begin{itemize}
\item
On the CFT side we can indeed modify the definition of a bulk field by adding something which has analytic correlators at spacelike separation.  But such a modification by definition respects
locality, i.e.\ it's a local field redefinition in the bulk.
\item
On the bulk side we are always free to make local field redefinitions.  But this changes the definition of a bulk field in a way that respects locality, i.e.\ the modification has correlators which are
analytic at spacelike separation.
\end{itemize}

We believe this ambiguity is sufficient to resolve the puzzles raised at the start of this section: to ${\cal O}(1/N)$, any cubic vertex in the bulk can be put in the form of a simple $\phi^3$ interaction
by a field redefinition.
In the rest of this section we study a few examples in more detail, to gain a better understanding of the ambiguity and its origins and implications on both the CFT and bulk sides.

\subsection{A bulk field redefinition}
Consider a bulk theory that has three scalar fields with a derivative coupling ${\lambda \over N} \phi_1 \partial_M \phi_2 \partial^M \phi_3$ and an equation of motion
\be
\left(\nabla^2 - m_1^2\right) \phi_1 = {\lambda \over N} \partial_M \phi_2 \partial^M \phi_3
\ee
With a superscript ${}^{(0)}$ denoting a free field,\footnote{From the bulk point of view this means zeroth order in perturbation theory.  Such a field can be constructed in the CFT
using the lowest-order smearing relation (\ref{eom}).} note that
\be
\nabla^2 \left(\phi_2^{(0)} \phi_3^{(0)}\right) = \left(m_2^2 + m_3^2\right) \phi_2^{(0)} \phi_3^{(0)} + 2 \partial_M \phi_2^{(0)} \partial^M \phi_3^{(0)}
\ee
This means that if we redefine the bulk field
\be
\tilde{\phi}_1 = \phi_1 - {\lambda \over 2N} \phi_2 \phi_3
\label{redef}
\ee
the new field will obey the equation of motion
\be
\left(\nabla^2 - m_1^2\right) \tilde{\phi}_1 = {\lambda \over 2N} \left(m_1^2 - m_2^2 - m_3^2\right) \phi_2 \phi_3 + {\cal O}(\lambda^2/N^2)
\ee
This shows that, to ${\cal O}(1/N)$, the derivative coupling we started from can be turned into a simple $\phi^3$ coupling by a field redefinition.  We believe this is true for any bulk interaction that
gives rise to a cubic coupling: to ${\cal O}(1/N)$ it is related to a $\phi^3$ coupling by a field redefinition.\footnote{Generically these interactions could be distinguished at ${\cal O}(1/N^2)$, that is, at the level of 4-point functions.}  This explains why we were always able to interpret the equations of motion we derived
from the CFT as corresponding to a $\phi^3$ coupling in the bulk.  It also means that the derivative coupling theory induces a boundary 3-point function which is $\frac{1}{2}(m_{1}^2-m_{2}^2-m_{3}^{2})$ times
larger than the regular cubic theory. This result was noted in \cite{Freedman:1998tz}.

It's clear that field redefinitions will change the bulk equations of motion.  It's interesting to ask how field redefinitions look when we think of bulk fields, not in terms of equations of motion,
but rather in terms of a sum over higher-dimension operators as in (\ref{basicbulkcft}).  Suppose we represent
\begin{equation}
\tilde{\phi}_{i}(x,z)=\int d^dx' K_{i}(x,z \vert x') {\cal O}_i(x') + {1 \over N} \sum_n \tilde{a}^{CFT}_{n} \int d^dx' K_{n}(x,z \vert x') {\cal O}_{n}(x')
\label{phitilderep}
\end{equation}
where the coefficients $\tilde{a}_n^{CFT}$ are chosen to make $\tilde{\phi}$ local in the bulk.  The product $\phi_2 \phi_3$ is a local scalar in the bulk, so as discussed in appendix \ref{appendix:bulk},
to ${\cal O}(1/N)$ it must be possible to write it as a sum of smeared primary double-trace operators.
\begin{equation}
\phi_{2}^{(0)}\phi_{3}^{(0)}(x,z) = \sum_{n} b_n \int d^dx' K_{\Delta_n}(x,z \vert x'){\cal O}_{n}(x')
\label{phi0phi0redef}
\end{equation}
This is identical to the representation used in (\ref{phi0phi0}).  From (\ref{redef}) this means the original scalar field $\phi_1$, which is derivatively coupled to $\phi_2$ and $\phi_3$, has a
representation involving a sum of higher-dimension operators just as in (\ref{phitilderep}), but with coefficients that are given by
\be
a_n = \tilde{a}_n + {\lambda \over 2} b_n
\ee
The correction ${\lambda \over 2} b_n$ is such that, after summing against an infinite tower of higher-dimension operators, it produces ${\lambda \over 2N}\phi_{2}^{(0)}\phi_{3}^{(0)}$ which to ${\cal O}(1/N)$
has analytic correlators at spacelike separation.  The correction therefore represents an ambiguity in our prescription for recovering bulk physics from the CFT.

\subsection{A CFT ambiguity}
It's clear that our prescription for building up bulk fields by requiring microcausality is ambiguous, because we're always free to add operators which have analytic correlators at spacelike separation.  But one could
ask where this freedom manifests itself in the procedure we followed, of summing the analytic parts of correlators to derive bulk equations of motion.

In fact there is a natural place where such ambiguities arise from the CFT point of view.
We have been using equations such as (\ref{fnu}) and (\ref{fnul}) to get a formula for the coefficients $a^{CFT}_{n}c_{njk}$. Formally these equations cannot be unique. The reason is as follows.  In the previous two subsections we have been using the property that all the functions which appear are eigenfunctions of the same differential operator $L_{0}$, just with different eigenvalues. So for example acting with
$L_{0}$ on (\ref{fnu}) gives, at least formally, a different summation formula.
\begin{equation}
f_{\nu}(x)=\frac{\cos(\frac{\pi \nu}{2})}{\nu(\nu+1)}\sum_{k}\frac{(4k+3)(2k+1)(2k+2)}{(2k+1-\nu)(2k+2+\nu)}P_{2k+1}(x)
\label{fnumod}
\end{equation}
Use of this formula will give different results for the coefficients $a_{n}^{CFT}$.  While the new coefficients will still cancel the non-analytic parts of the correlator they give a different summation leading to different equations
of motion. 

As an example take $d=1$ where we saw that the order of the Legendre function is $\Delta_{n}-1$, where $\Delta_{n}$ is the dimension of the operator which is smeared into the bulk.
Acting with $L_{0}$ gives
\begin{equation}
L_{0} P_{\Delta_{n}-1}=(\Delta_n -1)\Delta_n P_{\Delta_{i}-1}=M_{n}^2P_{\Delta_{i}-1}
\end{equation}
This means the difference between the coefficients coming from the use of (\ref{fnu}) and the coefficients coming from the modified summation formula (\ref{fnumod}) is 
\begin{equation}
a_{n}^{mod}=\frac{M_{n}^2}{m_{\nu}^2}a_{n}
\end{equation}
Putting this into the sum defining the bulk field one gets
\begin{equation}
\sum_{n=0}^{\infty}\frac{M_{n}^2}{m_{\nu}^2}a_{n}\int K_{\Delta_{n}}{\cal O}_{n}=\sum_{n=0}^{\infty}a_{n}\int K_{\Delta_{n}}{\cal O}_{n}+\frac{1}{m_{\nu}^{2}}\sum_{n=0}^{\infty}(M_{n}^2-m_{\nu}^{2})a_{n}\int K_{\Delta_{n}}{\cal O}_{n}
\end{equation}
From the equation of motion (\ref{Phi3eom}) it follows that $(M_{n}^{2}-m_{\nu}^{2})a_{n}=\lambda b_{n}$.  So we have $a_{n}^{mod}=a_{n}+\frac{\lambda}{m_{\nu}^{2}}b_{n}$, or in other words
\begin{equation}
\phi_{\nu}^{mod}(x,z)=\phi_{\nu}(x,z)+\frac{\lambda}{Nm_{\nu}^{2}}\phi^{0}_{j}\phi^{0}_{k}(x,z)
\end{equation}
This is just a local field redefinition.

Another option is to act with some $L_{i}$ on (\ref{fnu}) to eliminate one of the higher-dimension operators from the sum.  For example suppose we eliminate the operator which corresponds to a field of
mass squared $M_{i}^{2}$. Then one gets a modified definition of the bulk operator 
\begin{equation}
\sum_{n\neq i}^{\infty}\frac{M_{n}^2-M_{i}^{2}}{m_{\nu}^2-M_{i}^{2}}a_{n}\int K_{\Delta_{n}}{\cal O}_{n}=\sum_{n=0}^{\infty}a_{n}\int K_{\Delta_{n}}{\cal O}_{n}+\frac{1}{m_{\nu}^{2}-M_{i}^{2}}\sum_{n=0}^{\infty}(M_{n}^2-m_{\nu}^{2})a_{n}\int K_{\Delta_{n}}{\cal O}_{n}
\end{equation}
So $a_{n}^{mod}=a_{n}+\frac{\lambda}{m_{\nu}^{2}-M_{i}^{2}}b_{n}$.  This corresponds to a local field redefinition
\begin{equation}
\phi_{\nu}^{mod}(x,z)=\phi_{\nu}(x,z)+\frac{\lambda}{N}\frac{1}{m_{\nu}^{2}-M_{i}^{2}}\phi^{0}_{j}\phi^{0}_{k}(x,z)
\end{equation}

\section{CFT computation in $d=2$\label{sect:d=2}}
In this section we consider interacting scalar fields in AdS${}_3$.  We start from the CFT 3-point function of a scalar operator of dimension $\Delta_{\nu}$ (which we will promote to a
bulk field) and two operators of dimension $\Delta_j=\Delta_k=l$.
The discontinuity is given by (\ref{discont}) to be
\begin{equation}
I_{discont}^{\Delta_{\nu}}=\tilde{\gamma}^{(d=2)}_{\Delta_{\nu} ll}\frac{\Gamma(\Delta_{\nu})}{\Gamma^2(\frac{\Delta_{\nu}}{2})}F(\frac{\Delta_{\nu}}{2},1-\frac{\Delta_{\nu}}{2},1,1-\chi)
\end{equation}
Using $F(\nu+1,-\nu,1,1-\chi)=P_{\nu}(2\chi-1)$ we get 
\begin{equation}
I_{discont}^{\Delta_{\nu}}=\tilde{\gamma}^{(d=2)}_{\Delta_{\nu} ll}\frac{\Gamma(\Delta_{\nu})}{\Gamma^2(\frac{\Delta_{\nu}}{2})}P_{\frac{\Delta_{\nu}}{2}-1}(2\chi-1)
\label{deq2Discont}
\end{equation}
For a double-trace operator of dimension $2l+2n$ we get the discontinuity
\begin{equation}
\label{I2np2l}
I^{2n+2l}_{discont}=c_{nll}\frac{\Gamma(2n+2l)}{\Gamma^2(n+l)}P_{n+l-1}(2\chi-1)
\end{equation}
So we need to write $P_{\frac{\Delta_{\nu}}{2}-1}(x)$ as a sum of Legendre polynomials. This could be done by
\begin{equation}
P_{\frac{\Delta_{\nu}}{2}-1}(x)=\frac{4}{\pi} \sin(\pi\Delta_{\nu}/2) \sum_{k=0}^{\infty}(-1)^{k}\frac{2k+1}{(2k-\Delta_{\nu}+2)(2k+\Delta_{\nu})}P_{k}(x)
\label{basicd2}
\end{equation}
But in (\ref{I2np2l}) we only have available Legendre polynomials for $k \geq l-1$, so we want to use something like (\ref{basicd2}) but without the
first $l-1$ terms. This seems impossible, but can be achieved by acting with the operator (see (\ref{operator}))
\begin{equation}
\prod_{i=0}^{l-2}L_{i}
\end{equation}
on both sides of (\ref{basicd2}). This gives the formal expression
\begin{equation}
P_{\frac{\Delta_{\nu}}{2}-1}(x)=\frac{4}{c_{2,\nu,l}}\frac{\sin(\pi\Delta_{\nu}/2)}{\pi}\sum_{k=l-1}^{\infty}(-1)^{k}\frac{(2k+1)\prod_{i=0}^{l-2}[k(k+1)-i(i+1)]}{(2k-\Delta_{\nu}+2)(2k+\Delta_{\nu})}P_{k}(x)
\label{basicd2g}
\end{equation}
where
\be
c_{2,\nu,l} = \prod_{i=0}^{l-2}[(\frac{\Delta_{\nu}}{2}-1)(\frac{\Delta_{\nu}}{2})-i(i+1)]=(-1)^{l-1}\frac{\Gamma(l-\frac{\Delta_{\nu}}{2})\Gamma(\frac{\Delta_{\nu}}{2}+l-1)}{\Gamma(1-\frac{\Delta_{\nu}}{2})\Gamma(\frac{\Delta_{\nu}}{2})}
\ee
Also the product in (\ref{basicd2g}) can be evaluated as
\be
\prod_{i=0}^{l-2}[k(k+1)-i(i+1)]=\frac{\Gamma(k+l)}{\Gamma(k-l+2)}
\ee
We can now compute the coefficients we need for bulk locality. Setting $k=n+l$
and using $\Delta_n = 2n+2l$ we have
\begin{equation}
(2n+2l-\Delta_{\nu})(2n+2l+\Delta_{\nu}-2)=\Delta_{n}(\Delta_{n}-2)-\Delta_{\nu}(\Delta_{\nu}-2)
\end{equation}
The condition for canceling the discontinuity (\ref{basiccondition}) gives
\begin{eqnarray}
& &a_{n}^{CFT}c_{nll}=-N\tilde{\gamma}^{(d=2)}_{\Delta_{\nu} ll}\frac{4\Gamma(\Delta_{\nu})}{\Gamma^2(\frac{\Delta_{\nu}}{2})\Gamma(l-\frac{\Delta_{\nu}}{2})\Gamma(l+\frac{\Delta_{\nu}}{2}-1)} \nonumber\\
& & \times \frac{\Gamma^{2}(n+l)\Gamma(n+2l-1)}{\Gamma(n+1)\Gamma(2n+2l-1)}\frac{(-1)^{n}}{\Delta_{n}(\Delta_{n}-2)-\Delta_{\nu}(\Delta_{\nu}-2)}
\label{cftcoefd2}
\end{eqnarray}
Now from (\ref{corgijk})
\begin{equation}
\tilde{\gamma}_{\Delta_{\nu} ll}^{(d=2)}=-\frac{\lambda}{4\pi^2 N}\frac{\Gamma^2(\frac{\Delta_{\nu}}{2})\Gamma(l-\frac{\Delta_{\nu}}{2})\Gamma(l+\frac{\Delta_{\nu}}{2}-1)}{\Gamma(\Delta_{\nu})\Gamma^{2}(l-1)}
\end{equation}
Putting this together, again we find agreement with the general result (\ref{coefbulk}).

\subsection{Even $\Delta_{\nu}$}
The preceding formulas do not apply when $\Delta_{\nu}$ is even, so we work out an explicit example to illustrate this case. We consider $\Delta_{\nu}=\Delta_{j}=\Delta_{k}=2$ which describes a massless scalar field in AdS${}_{3}$.

The discontinuity we need to cancel is given by
\begin{equation}
I_{discont}^{2}=\tilde{\gamma}^{(d=2)}_{222}P_{0}(2\chi-1)
\end{equation}
with $\tilde{\gamma}^{(d=2)}_{222}=-\frac{\lambda}{4N\pi^2}$.
The double-trace operators we have available have dimensions $\Delta_n = 2n + 4$, and their contribution to the discontinuity
is given by
\begin{equation}
I^{2n+4}_{discont}=c_{n22}\frac{\Gamma(2n+4)}{\Gamma^{2}(n+2)}P_{n+1}(2\chi-1)
\end{equation}

To obtain an identity that will let us cancel the discontinuities we start with (\ref{basicd2}) and act on both sides with the operator $L_{0}$.
This gives the formal expression
\begin{equation}
\frac{\pi\frac{\Delta_{\nu}}{2}(\frac{\Delta_{\nu}}{2}-1)}{\sin\pi(\frac{\Delta_{\nu}}{2})}P_{\frac{\Delta_{\nu}}{2}-1}(x)=4\sum_{k=0}^{\infty}(-1)^{k}\frac{k(k+1)(2k+1)}{(2k-\Delta_{\nu}+2)(2k+\Delta_{\nu})}P_{k}(x)
\end{equation}
Now we send $\Delta_{\nu} \rightarrow 2$ on both sides to obtain
\begin{equation}
P_{0}(x)=-\sum_{k=1}^{\infty}(-1)^k(2k+1)P_{k}(x)
\end{equation}
This means (\ref{basiccondition}) can be solved by taking the coefficients to be
\begin{equation}
a_{n}^{CFT}c_{n22}=\frac{\lambda}{4\pi^2}(-1)^n \frac{\Gamma^2(n+2)}{\Gamma(2n+3)}
\end{equation}
Using $(2n+4)(2n+2)-2=4(n+1)(n+2)$, we get
\begin{equation}
a_{n}^{CFT}c_{n22}=\lambda \frac{(-1)^n}{\pi^2}\frac{\Gamma^2(n+2)\Gamma(n+3)}{\Gamma(n+1)\Gamma(2n+3)}\frac{1}{(2n+4)(2n+2)-2}
\end{equation}
which again agrees with the general result (\ref{coefbulk}).

\subsection{Equations of motion\label{EvendEOM}}
Now let's study the equation of motion (\ref{eomcft}) that our local bulk field satisfies.  The basic strategy is the same as in section \ref{sect:bulkeom}, except that
in even dimensions we cannot use (\ref{anapartd}), so we will have to proceed by a slightly different route.

For the case $\Delta_{j}=\Delta_{k}$ the double-trace operators we can construct have dimension $\Delta_n = 2\Delta_j+2n$.  From (\ref{3pointbulk}) the
3-point function of a smeared ${\cal O}_n$ with ${\cal O}_j$ and ${\cal O}_k$ is
\begin{equation}
\frac{c_{njk}}{(y_1-y_2)^{2\Delta_{j}}}\frac{1}{(\chi-1)^{\DS}}F(\DS,\DS,2\DS,\frac{1}{1-\chi})
\label{anald2}
\end{equation}
Here $\DS = \Delta_n / 2$.
We now use a hypergeometric identity
\begin{equation}
F(a,b,2b,z)=\big(1-\frac{z}{2}\big)^{-a}F\big(\frac{a}{2},\frac{a+1}{2},b+\frac{1}{2},(\frac{z}{2-z})^2\big)
\end{equation}
followed by the transformation (\ref{hypertrans1}). We have to be a little careful with what branch we are on so in the second transformation we introduce
a small $\pm i\epsilon$ term.  Then (\ref{anald2}) becomes
\begin{equation}
\frac{c_{njk}}{(y_1-y_2)^{2\Delta_{j}}}\frac{\pi (-2)^{\DS}\Gamma(\DS+\frac{1}{2})}{\Gamma(\frac{\DS+1}{2})\Gamma(\frac{\DS}{2})}\big[\frac{2}{\pi}Q_{\DS-1}(1-2\chi)+(-1\pm i\epsilon)^{-\frac{1}{2}}P_{\DS-1}(1-2\chi)\big]
\label{analyticd2part}
\end{equation}
As in (\ref{legendre1}) the Legendre function $Q_\nu(x)$ is defined to have branch cuts
along $(-\infty,-1) \cup (1,\infty)$, so the first term in (\ref{analyticd2part}) does not have a cut for $0 < \chi < 1$.
The unwanted discontinuity across the cut (\ref{deq2Discont}) comes just from the second term.
When summing the tower of higher-dimension operators to obtain the equation of motion, the coefficients
are chosen so the unwanted discontinuity cancels.  This means only the first term in (\ref{analyticd2part}) contributes to the equations of motion.

In deriving equations of motion we first treat the case $\Delta_j=2$. From the coefficients we have computed we have
\begin{equation}
(M_{n}^2-\Delta_{\nu}(\Delta_{\nu}-2))a^{CFT}_{n}c^{jk}_{n}=\lambda\frac{(-1)^n}{\pi^2}\frac{\Gamma^2(n+2)\Gamma(n+3)}{\Gamma(n+1)\Gamma(2n+3)}
\end{equation}
With the repeated use of 
\begin{equation}
\Gamma(2x)=\frac{2^{2x-1}}{\sqrt{\pi}}\Gamma(x)\Gamma(x+\frac{1}{2})
\end{equation}
the sum we need to do on the right hand side of (\ref{eomcft}) reduces to
\begin{equation}
\frac{2\lambda}{\pi^2(y_1-y_2)^{4}}\sum_{n=0}^{\infty}(n+1)(n+2)(2n+3)Q_{n+1}(1-2\chi).
\label{anasum2}
\end{equation}
To evaluate this sum we start with
\begin{equation}
\sum_{k=0}^{\infty}(2k+1)Q_{k}(x)=\frac{1}{x-1}
\label{basicsumd2}
\end{equation}
and act with $L_{0}$ on both sides to get
\begin{equation}
\sum_{n=0}^{\infty}(2n+3)(n+1)(n+2)Q_{n+1}(x)=L_{0}\frac{1}{x-1}=\frac{2}{(1-x)^2}
\end{equation}
Thus we get that the sum (\ref{anasum2}) is just
\begin{equation}
\frac{\lambda}{\pi^2}\left(\frac{z}{(x-y_1)^2 +z^2}\right)^2\left(\frac{z}{(x-y_2)^2 +z^2}\right)^2
\end{equation}
According to (\ref{Maincoef2}) and (\ref{Maindjk}) this means the sum over double-trace operators on the right hand side of the equation of motion
can be identified with $\lambda \phi^{(0)}_{j}\phi^{(0)}_{k}(x,z)$.

Now we consider the more general case $\Delta_{j}=\Delta_{k}=l\geq2$. Using (\ref{cftcoefd2}) and (\ref{analyticd2part}) and a little algebra we need to do the sum
\begin{equation}
\frac{\lambda}{(y_1-y_2)^{2l}}\frac{2(-1)^{l}}{\pi^2 \Gamma^{2}(l-1)}\sum_{n=0}^{\infty}\frac{(2n+2l-1)\Gamma(n+2l-1)}{\Gamma(n+1)}Q_{n+l-1}(1-2\chi)
\label{gensumd2}
\end{equation}
To evaluate this we start with (\ref{basicsumd2}) and eliminate the first $l-1$ terms by acting with $L_{l-2}\cdots L_{i-2} \cdots L_{0}$.
Using $L_{i}Q_{k}=(k-i)(k+i+1)Q_{k}$ we find
\begin{equation}
L_{l-2}\cdots L_{i-2} \cdots L_{0}\sum_{k=0}^{\infty}(2k+1)Q_{k}(x)=\sum_{n=0}^{\infty}\frac{(2n+2l-1)\Gamma(n+2l-1)}{\Gamma(n+1)}Q_{n+l-1}(x)
\end{equation}
A bit of computation gives
\begin{equation}
L_{i}\frac{1}{(x-1)^{i+1}}=\frac{2(i+1)^2}{(x-1)^{i+2}}
\end{equation}
so that
\begin{equation}
L_{l-2}\cdots L_{i-2} \cdots L_{0}\frac{1}{x-1}=\frac{2^{l-1}\Gamma^{2}(l)}{(x-1)^l}.
\end{equation}
This means the sum in (\ref{gensumd2}) is given by
\begin{equation}
\frac{\lambda}{(y_1-y_2)^{2l}}\frac{\Gamma^{2}(l)}{\pi^2\Gamma^{2}(l-1)}\frac{1}{\chi^l}=\frac{\lambda \Gamma^{2}(l)}{\pi^2\Gamma^{2}(l-1)}\left(\frac{z}{(x-y_1)^2+z^2}\right)^{l}\left(\frac{z}{(x-y_2)^2+z^2}\right)^{l}
\end{equation}
Again using (\ref{Maincoef2}) and (\ref{Maindjk}) the sum over double-trace operators on the right hand side of the equation of motion can be identified
with $\lambda \phi^{(0)}_{j}\phi^{(0)}_{k}(x,z)$.
This means the CFT expression (\ref{eomcft}) for the bulk equation of motion is equivalent to
\begin{equation}
(\nabla^2-m^2)\phi(x,z)=\frac{\lambda}{N}\phi_{j}^{(0)}\phi_{k}^{(0)}(x,z)
\end{equation}

\section{CFT computation in $d=3$\label{sect:d=3}}
Now we consider scalar fields in AdS${}_4$.
We start with the 3-point function of a smeared dimension $\Delta_{\nu}$ operator with two operators of dimension $\Delta_j=\Delta_k=l \in {\mathbb N}$.
In AdS${}_4$ it is convenient to define $\Delta_\nu = \nu + 2$.

The discontinuity (\ref{discont}) is
\begin{equation}
I_{discont}^{\Delta_{\nu}}=\tilde{\gamma}_{\Delta_{\nu}ll}^{(d=3)}\frac{1}{(1-\chi)^{\frac{1}{2}}}\frac{\Gamma(\Delta_{i}-\frac{1}{2})}{\sqrt{\pi}\Gamma^{2}(\frac{\Delta_{i}}{2})}
F(\frac{\Delta_{\nu}}{2}-\frac{1}{2},1-\frac{\Delta_{i}}{2},\frac{1}{2},1-\chi)
\end{equation}
Using (\ref{legendre1}) we can write for generic $\nu$
\begin{eqnarray}
&& F(\frac{\nu+1}{2},-\frac{\nu}{2},\frac{1}{2},x^2)=\frac{\Gamma(\frac{\nu}{2}+1)}{\sqrt{\pi}\Gamma(\frac{\nu+1}{2})}g_{\nu}(x)\nonumber\\
&& g_{\nu}(x)=\left(\pi\cos(\frac{\pi \nu}{2})P_{\nu}(x)-2\sin(\frac{\pi \nu}{2})Q_{\nu}(x)\right)
\end{eqnarray}
The double-trace operators we have available start at $\Delta_{n}=2l+2n$ and produce a discontinuity given by
\begin{equation}
I_{discont}^{2n+2l}=\frac{c_{njk}}{(1-\chi)^{\frac{1}{2}}}\frac{\Gamma(2n+2l-\frac{1}{2})}{\Gamma(n+l)\Gamma(n+l-\frac{1}{2})}(-1)^{n+l-1}P_{2n+2l-2}(\sqrt{1-\chi})
\label{dtd3}
\end{equation}

Now let's see how we can cancel these discontinuities.  From the integrals in (\ref{legendre2}) we can derive the identity
\begin{equation}
g_{\nu}(x)=-2\sin(\frac{\pi \nu}{2})\sum_{k=0}^{\infty}\frac{4k+1}{(2k-\nu)(2k+\nu+1)}P_{2k}(x)
\label{gnu}
\end{equation}
However for $l \geq 2$ we do not have all the required Legendre polynomials in (\ref{dtd3}).  So as in the previous section we act on both sides of
(\ref{gnu}) with the operator (see (\ref{operator}))
\begin{equation}
\Pi_{i=0}^{l-2}L_{2i}
\end{equation}
to get the formal expression
\begin{equation}
g_{\nu}(x)=\frac{-2\sin(\frac{\pi \nu}{2})}{h_{3,\nu,l}}\sum_{n=0}^{\infty}\frac{(4n+4l-3)\Pi_{i=0}^{l-2}[(2n+2l-2)(2n+2l-1)-2i(2i+1)]}{(2n+2l-2-\nu)(2n+2l+\nu-1)}P_{2n+2l-2}(x)
\label{gnug}
\end{equation}
Here
\begin{equation}
h_{3,\nu,l}=\Pi_{i=0}^{l-2}[\nu(\nu+1)-2i(2i+1)]=2^{2l-2}(-1)^{l-1}\frac{\Gamma(l-\frac{\nu}{2}-1)\Gamma(l+\frac{\nu}{2}-\frac{1}{2})}{\Gamma(-\frac{\nu}{2})\Gamma(\frac{\nu+1}{2})}
\end{equation}
The product in (\ref{gnug}) evaluates to
\begin{equation}
\Pi_{i=0}^{l-2}[(2n+2l-2)(2n+2l-1)-2i(2i+1)]=2^{2l-2}\frac{\Gamma(n+2l-\frac{3}{2})\Gamma(n+l)}{\Gamma(n+1)\Gamma(n+l-\frac{1}{2})}
\end{equation}
It's also useful to use $2n+2l=\Delta_{n}$ and $\nu=\Delta_{\nu}-2$ to rewrite
\begin{equation}
(2n+2l-2-\nu)(2n+2l+\nu-1)=\Delta_{n}(\Delta_{n}-3)-\Delta_{\nu}(\Delta_{\nu}-3)
\end{equation}
Now let's cancel the discontinuities as in (\ref{basiccondition}).  We need to cancel 
\begin{equation}
\tilde{\gamma}^{(d=3)}_{\Delta_{\nu} ll}\frac{1}{(1-\chi)^{1/2}}\frac{\Gamma(\Delta_{\nu}-\frac{1}{2})}{\pi\Gamma(\frac{\Delta_{\nu}}{2})\Gamma(\frac{\Delta_{\nu}-1}{2})}g_{\nu}(\sqrt{1-\chi})
\end{equation}
against terms of the form (\ref{dtd3}).  This can be done by setting
\begin{eqnarray}
& &a_{n}^{CFT}c_{njk}=-N\tilde{\gamma}^{(d=3)}_{\Delta_{\nu} ll}\frac{4}{\pi}\frac{\sin(\frac{\pi \nu}{2})\Gamma(\nu+\frac{3}{2})\Gamma(-\frac{\nu}{2})}{\Gamma(\frac{\nu+2}{2})\Gamma(l-\frac{\nu}{2}-1)\Gamma(l+\frac{\nu}{2}-\frac{1}{2})}\times \nonumber\\
& &\frac{\Gamma(n+2l-\frac{3}{2})\Gamma^{2}(n+l)}{\Gamma(n+1)\Gamma(2n+2l-\frac{3}{2})}\frac{(-1)^{n}}{\Delta_{n}(\Delta_{n}-3)-\Delta_{\nu}(\Delta_{\nu}-3)}
\label{coefd3}
\end{eqnarray}
Now
\begin{equation}
\tilde{\gamma}^{(d=3)}_{\Delta_{\nu} ll}=-\frac{\lambda}{4\pi^3 N}\frac{\Gamma^{2}(\frac{\nu}{2}+1)\Gamma(l-\frac{\nu}{2}-1)\Gamma(l+\frac{\nu}{2}-\frac{1}{2})}{\Gamma(\nu +\frac{3}{2})\Gamma^2(l-\frac{3}{2})}
\end{equation}
Putting everything together one again finds that the coefficients agree with the general result (\ref{coefbulk}).

\subsection{Equations of motion}
As an example of deriving equations of motion in $d=3$ we consider operators of dimension $\Delta_j = \Delta_k = l$ coupled to an operator with
dimension $\Delta_{\nu}$. Dropping non-analytic terms which will cancel anyway, according to (\ref{eomcft}) and (\ref{anapartd}) the terms we
need to sum are
\begin{equation}
c_{njk}\frac{\Gamma(-\frac{1}{2})\Gamma(2n+2l-\frac{1}{2})}{(y_{1}-y_{2})^{2l}\Gamma^2(n+l-\frac{1}{2})}F(n+l,-n-l+\frac{3}{2},\frac{3}{2},1-\chi)
\label{basicsumd3-1}
\end{equation}
Now
\begin{equation}
F(n+l,-n-l+\frac{3}{2},\frac{3}{2},1-\chi)
=\frac{(-1)^{n+l-1}\Gamma(n+l-\frac{1}{2})}{\sqrt{\pi}\Gamma(n+l)}\frac{1}{\sqrt{1-\chi}}Q_{2n+2l-2}(\sqrt{1-\chi})
\label{basicsumd3-2}
\end{equation}
while from (\ref{coefd3}) we have
\begin{equation}
[\Delta_{n}(\Delta_{n}-3)-\Delta_{\nu}(\Delta_{\nu}-3)]a^{CFT}c_{n}=\lambda
\frac{(-1)^n}{\pi^3 \Gamma^{2}(l-\frac{3}{2})}\frac{\Gamma^{2}(n+l)\Gamma(n+2l-\frac{3}{2})}{\Gamma(n+1)\Gamma(2n+2l-\frac{3}{2})}
\label{basicsumd3-3}
\end{equation}
As a first example we consider the case $l=2$.  From (\ref{basicsumd3-1}), (\ref{basicsumd3-2}), (\ref{basicsumd3-3}) we need to do the sum
\begin{equation}
\frac{\lambda}{N(y_1-y_2)^4}\frac{1}{4\pi^4}\frac{1}{\sqrt{1-\chi}}\sum_{n=0}^{\infty}(2n+2)(2n+3)(4n+5)Q_{2n+2}(\sqrt{1-\chi})
\label{sumd31}
\end{equation}
We can use (\ref{basican}) to get
\begin{equation}
\sum_{k=0}^{\infty}(4k+1)Q_{2k}(x)=\frac{x}{(x^2-1)}
\label{basicand3}
\end{equation}
But since we only have $Q_{2k}(x)$ available for $k\geq 1$, we act on (\ref{basicand3}) with $L_{0}$ to eliminate $Q_{0}(x)$.  This gives
a formal expression
\begin{equation}
\sum_{n=0}^{\infty}(4n+5)(2n+2)(2n+3)Q_{2n+2}(x)=\frac{4x}{(x^2-1)^2}
\end{equation}
which means the sum (\ref{sumd31}) is
\begin{equation}
\frac{\lambda}{\pi^4}\left(\frac{z}{(x-y_1)^2+z^2}\right)^{2}\left(\frac{z}{(x-y_2)^2+z^2}\right)^{2}
\end{equation}
Using (\ref{Maincoef2}) and (\ref{Maindjk}), this identifies the right hand side of the equation of motion (\ref{eomcft}) as simply being
$\frac{\lambda}{N} \phi^{(0)}_{j}\phi^{(0)}_{k}(x,z)$.

Now we consider general $l > 2$. From (\ref{basicsumd3-1}), (\ref{basicsumd3-2}), (\ref{basicsumd3-3}) we need to evaluate
\begin{equation}
\frac{\lambda}{(y_1-y_2)^{2l}}\frac{(-1)^{l}}{\pi^3 \Gamma^{2}(l-\frac{3}{2})}\frac{1}{\sqrt{1-\chi}}\sum_{n=0}^{\infty}(4n+4l-3)\frac{\Gamma(n+l)\Gamma(n+2l-\frac{3}{2})}{\Gamma(n+1)\Gamma(n+l-\frac{1}{2})}Q_{2n+2l-2}(\sqrt{1-\chi})
\label{totsumd3}
\end{equation}
Starting from (\ref{basicand3}) we need to eliminate the first $l-1$ terms, which we do by acting with $L_{2l-4} \cdots L_0$. Using $L_{i}Q_{k}=(k-i)(k+i+1)Q_{k}$
the left hand side becomes
\begin{eqnarray}
& &L_{2l-4}\cdots L_{2i-4}\cdots L_{0}\sum_{k=0}^{\infty}(4k+1)Q_{2k}(x)=\nonumber\\
& &4^{l-1}\sum_{n=0}^{\infty}(4n+4l-3)\frac{\Gamma(n+l)\Gamma(n+2l-\frac{3}{2})}{\Gamma(n+1)\Gamma(n+l-\frac{1}{2})}Q_{2n+2l-2}(\sqrt{1-\chi})
\end{eqnarray}
Using
\begin{equation}
L_{2i-4}\frac{x}{(x^2-1)^{i-1}} =4(i-1)^2\frac{x}{(x^2-1)^{i}}
\end{equation}
the right hand side becomes
\begin{equation}
L_{2l-4}\cdots L_{2i-4}\cdots L_{0}\frac{x}{x^2-1}=4^{l-1}\Gamma^{2}(l)\frac{x}{(x^2-1)^l}
\end{equation}
Putting everything together the sum in (\ref{totsumd3}) is
\begin{equation}
\frac{\lambda}{(y_1-y_2)^{2l}}\frac{\Gamma^{2}(l)}{\pi^3 \Gamma^{2}(l-\frac{3}{2})}\frac{1}{\chi^{l}}=\frac{\lambda\Gamma^{2}(l)}{\pi^3 \Gamma^{2}(l-\frac{3}{2})}\left(\frac{z}{(x-y_1)^2+z^2}\right)^{l}\left(\frac{z}{(x-y_2)^2+z^2}\right)^{l}
\end{equation}
Again this identifies the right hand side of the equation of motion (\ref{eomcft}) as $\frac{\lambda}{N} \phi^{(0)}_{j}\phi^{(0)}_{k}(x,z)$.

\section{Scalars coupled to gauge fields\label{sect:gauge}}
In this section we show that similar computations can be done for a charged scalar primary that has a non-trivial three-point function with a conserved current. We
promote the charged scalar primary to a local bulk field and derive the equation of motion, obtaining the expected bulk three-point coupling between a gauge field
and a charged scalar.

We start with a review of results from \cite{Kabat:2012av, Kabat:2013wga}.
We start with the 3-point function of a complex scalar and a conserved current in a CFT.
\begin{equation}
\langle{\cal O}(x) \bar{{\cal O}}(y_1) j_{\nu}(y_2)\rangle = \frac{-2i(q/N)\tilde{\xi} (d-2)}{(x-y_1)^{2\Delta-d+2}(x-y_{2})^{d-2}(y_1-y_2)^{d-2}}\left[\frac{(x-y_2)_{\nu}}{(x-y_2)^2}-\frac{(y_1-y_2)_{\nu}}{(y_1-y_2)^2}\right]
\end{equation}
Here $q = {\cal O}(1)$ is the charge of the field measured in units of $1/N$ and \cite{Freedman:1998tz}
\begin{equation}
\tilde{\xi}=\frac{\xi}{2\Delta-d}=\frac{\Gamma(\frac{d}{2})\Gamma(\Delta)}{2\pi^{d}(d-2)\Gamma(\Delta-\frac{d}{2})}\,.
\end{equation}
Since we have a charged field in the bulk, the Gauss constraint requires that certain commutators be non-vanishing even at spacelike separation.
To deal with this we follow a procedure developed in \cite{Kabat:2012av}, where it was shown that even taking the Gauss constraint into account,
a charged bulk scalar still commutes with the boundary field strength $F_{\mu\nu}=\partial_{\mu}j_{\nu}-\partial_{\nu}j_{\mu}$.
The relevant CFT correlator is given by
\begin{equation}
\langle{\cal O}(x) \bar{{\cal O}}(y_1) F_{\mu\nu}(y_2)\rangle =\frac{4iq}{N}(d-2) \tilde{\xi} L_{\mu\nu} \langle\Delta, \Delta+1, d-1\rangle 
\label{vec1}
\end{equation}
where
\begin{equation}
\langle\Delta_1, \Delta_2, \Delta_3\rangle =\frac{1}{(x-y_1)^{\Delta_1+\Delta_2-\Delta_3}(x-y_2)^{\Delta_1+\Delta_3-\Delta_2}(y_1-y_2)^{\Delta_2+\Delta_3-\Delta_1}}
\label{D1D2D3def}
\end{equation}
has a form of a scalar 3-point function with the indicated dimensions and
\begin{equation}
L_{\mu \nu}=(y_2-y_1)_{\mu}\partial_{\nu}^{y_2}-(\mu \leftrightarrow \nu)
\end{equation}
It was shown in \cite{Kabat:2012av, Kabat:2013wga} that one can construct a local bulk field, in the sense that it obeys the appropriate microcausality conditions, by setting
\begin{equation}
\phi(x,z) = \int K_{\Delta}(x,z \vert x') {\cal O}(x') + {1 \over N} \sum_{n=0}^{\infty}\int a^{(f)}_{n} K_{\Delta_{n}}(x,z \vert x' ){\cal A}_{n}(x').
\end{equation}
Here the scalar (but not primary!) operators ${\cal A}_{n}$ are built from $j_{\nu}$ and ${\cal O}$ with derivatives.  They have the property that
\begin{equation}
\langle{\cal A}_{n}(x) \bar{{\cal O}}(y_1) F_{\mu\nu}(y_2)\rangle =c^{(f)}_{n} L_{\mu \nu} \langle\Delta_{n}, \Delta+1, d-1\rangle 
\label{vec1a}
\end{equation}
where $\Delta_{n}=2n+\Delta+d$.
The coefficients $a_{n}^{(f)}$ are to be chosen such that 
\begin{equation}
\langle\phi(x,z) {\cal O}(y_1) F_{\mu\nu}(y_2)\rangle 
\end{equation}
is analytic for $\chi>0$.

The $a_{n}^{(f)}$ coefficients are related by (\ref{vec1}) and (\ref{vec1a})  to the coefficients one uses for interacting scalars with dimensions $\Delta, \Delta+1, d-1$. 
Suppose we solve the scalar case in the sense that we find coefficients $a^{(s)}_{n}c^{(s)}_{n}$ that would let us build a local scalar field in the bulk starting
from a CFT three-point coupling $\langle\Delta, \Delta+1, d-1\rangle$. This scalar system is solved explicitly for $d=3$ in appendix \ref{appendix:scalargauge}.
Then to build a charged bulk scalar field all we have to do is set
\begin{equation}
a^{(f)}_{n} c^{(f)}_{n}=\alpha^{(f)} a^{(s)}_{n}c^{(s)}_{n}
\label{svrel}
\end{equation}
where $\alpha^{(f)}$ reflects the different normalization of the starting 3-point functions.
The coefficient of the scalar 3-point function coming from a bulk cubic coupling is given in (\ref{corgijk}).
\begin{equation}
\tilde{\gamma}=-\frac{\lambda\Gamma(\frac{d}{2})\Gamma(\Delta)}{4N\pi^{d}\Gamma(\Delta-\frac{d}{2}+1)}
\end{equation}
Then we identify
\begin{equation}
\alpha^{(f)}=-4i\frac{q}{\lambda}(2\Delta-d)
\end{equation}

The resulting local bulk field obeys the equation of motion
\begin{equation}
(\nabla^2-\Delta(\Delta-d))\phi(x,z)= \frac{1}{N}\sum_{n=0}^{\infty}\int dx' a^{(f)}_{n}[\Delta_{n}(\Delta_{n}-d)-\Delta(\Delta-d)] K_{\Delta_{n}}(x,z \vert x' ){\cal A}_{n}(x')
\label{eomf}
\end{equation} 
To identify the operator on the right hand side we insert it into a 3-point function with $F_{\mu \nu}$ and $\bar{{\cal O}}$.  This gives
\begin{equation}
L_{\mu \nu} \sum_{n=0}^{\infty}\int dx' a^{(f)}_{n}c^{(f)}_{n}[\Delta_{n}(\Delta_{n}-d)-\Delta(\Delta-d)]K_{\Delta_{n}}(x,z \vert x' ) \langle\Delta_{n},\Delta+1, d-1\rangle 
\label{eomsumv}
\end{equation}
which using (\ref{svrel}) is just $\alpha^{(f)} L_{\mu \nu}$ acting on the scalar result given in appendix \ref{appendix:scalargauge}.
This means the right hand side of the equation of motion is
\begin{equation}
\label{ChargedRHS}
\alpha^{(f)}\frac{\lambda \Gamma(\Delta+1)}{\pi^3 \Gamma(\Delta-\frac{1}{2})\Gamma(\frac{1}{2})}L_{\mu \nu}\left(\frac{z}{z^2+(x-y_1)^2}\right)^{\Delta+1}\left(\frac{z}{z^2+(x-y_2)^2}\right)^{2}
\end{equation}

Now let's see if we can identify a bulk quantity which, when inserted in a 3-point function with $\bar{\cal O}(y_1)$ and $F_{\mu\nu}(y_2)$, will reproduce
(\ref{ChargedRHS}).  It's easy to guess that the answer must be $-2iqg^{\rho\sigma}A^{(0)}_{\rho}\partial_{\sigma}\phi^{(0)}(x,z)$, where to leading order in
$1/N$ $g^{\rho\sigma}$ is the background AdS metric and the superscript ${}^{(0)}$ indicates a free field constructed from the CFT using the zeroth order
smearing (same notation as in (\ref{eom})).

To check this we use large-$N$ factorization to obtain
\begin{equation}
<g^{\rho\sigma}A^{(0)}_{\rho}\partial_{\sigma}\phi^{(0)}(x,z)\bar{\cal O}(y_1)F_{\mu\nu}(y_2)\rangle =\langle zA^{(0)}_{\rho}(x,z)F_{\mu\nu}(y_2)\rangle \langle z\eta^{\rho\sigma}\partial_{\sigma}\phi^{(0)}(x,z)\bar{\cal O}(y_1)\rangle 
\end{equation}
Now we use \cite{Freedman:1998tz, Kabat:2012hp}
\begin{equation}
\langle zA^{(0)}_{\rho}(x,z)F_{\mu\nu}(y_2)\rangle =\frac{\Gamma(d-1)}{\pi^{\frac{d}{2}}\Gamma(\frac{d}{2}-1)}(\eta_{\rho\nu}\partial^{y_2}_{\mu}-\eta_{\rho\mu}\partial^{y_2}_{\nu})\left(\frac{z}{z^2+(x-y_2)^2}\right)^{d-1}
\label{scgauge2}
\end{equation}
and (note that for a complex scalar there is a factor of $2$ in the two-point function)
\begin{equation} 
\langle z\partial_{\sigma}\phi^{(0)}(x,z)\bar{\cal O}(y_1)\rangle =4(y_{1}-x)_{\sigma} \frac{\Gamma(\Delta+1)}{\pi^{\frac{d}{2}}\Gamma(\Delta-\frac{d}{2})} \left(\frac{z}{z^2 +(x-y_1)^2}\right)^{\Delta+1}
\label{scgauge3}
\end{equation}
We can replace $(y_1-x)_{\sigma}\rightarrow (y_1-y_2)_{\sigma}$ in (\ref{scgauge3}) due to the antisymmetry in (\ref{scgauge2}), and  we get  as promised  (setting  $d=3$)
\begin{eqnarray}
& &-2iq\langle g^{\rho\sigma}A^{(0)}_{\rho}\partial_{\sigma}\phi^{(0)}(x,z)\bar{\cal O}(y_1)F_{\mu\nu}(y_2)\rangle \\
& &= -8iq\frac{\Gamma(\Delta+1)}{\pi^3 \Gamma(\Delta-\frac{3}{2})\Gamma(\frac{1}{2})} L_{\mu\nu}\left(\frac{z}{z^2+(x-y_1)^2}\right)^{\Delta+1}\left(\frac{z}{z^2+(x-y_2)^2}\right)^{2}\nonumber
\end{eqnarray}
So the equation of motion (\ref{eomf}) is nothing but
\begin{equation}
 (\nabla^2-\Delta(\Delta-d))\phi(x,z)= -\frac{2iq}{N} g^{\rho\sigma}A^{(0)}_{\rho}\partial_{\sigma}\phi^{(0)}(x,z)
 \label{eomf1}
 \end{equation} 
 which is the expected bulk equation of motion for a charged scalar field in holographic gauge.
 
\subsection{Scalar coupled to a massive vector}
Another case that can be treated is a non-conserved current of dimension $\Delta$ coupled to two primary scalars of dimension $\Delta_{1}$ and $\Delta_{2}$.
The CFT 3-point function is given by
\begin{eqnarray}
\langle {\cal O}_{1}(x) j_{\mu}(y_1){\cal O}_{2}(y_2)\rangle &=& \frac{c^{(v)}_{\Delta_1}}{(x-y_1)^{\Delta+\Delta_{1}-\Delta_{2}-1}(x-y_{2})^{\Delta_1+\Delta_2-\Delta+1}(y_2-y_1)^{\Delta+\Delta_2-\Delta_1-1}}\nonumber\\
& &\times\left[\frac{(x-y_1)_{\mu}}{(x-y_1)^2}-\frac{(y_2-y_1)_{\mu}}{(y_2-y_1)^2}\right]
\end{eqnarray}
This can be written as
\begin{equation}
\frac{1}{(\Delta+\Delta_{1}-\Delta_{2}+1)}
[(y_2-y_1)^2\partial^{y_1}_{\mu}-2\Delta(y_2-y_1)_{\mu}]\langle \Delta_{1}, \Delta, \Delta_{2}+1\rangle \label{massive1}
\end{equation}
As explained in \cite{Kabat:2012av} we can build a tower of higher-dimension double-trace primary scalars ${\cal O}^{(v)}_{n}$ with dimension $\Delta_{n}=\Delta+\Delta_2+1 + 2n$ out of a product of $j_{\mu}$ and ${\cal O}_{2}$ with $2n$ derivatives. The operators ${\cal O}^{(v)}_{n}$ obey
\begin{equation}
\langle {\cal O}^{(v)}_{n}(x)j_{\mu}(y_1){\cal O}_{2}(y_2)\rangle =c^{(v)}_{n} [(y_2-y_1)^2\partial^{y_1}_{\mu}-2\Delta(y_2-y_1)_{\mu}]\langle \Delta_{n}, \Delta, \Delta_{2}+1\rangle 
\label{massive2}
\end{equation}

We see that this system is closely related to a system of interacting scalars of dimension $\Delta_1, \Delta, \Delta_{2}+1$. 
Suppose we solve the corresponding scalar system, that is we find a set of coefficients $a_{n}^{(s)}$ that produce a local scalar field in the bulk.
Then from (\ref{massive1}) and (\ref{massive2}), we see that we should define
\begin{equation}
\phi_{1}(x,z)=\int K_{\Delta_{1}}(x,z \vert x') {\cal O}_{1}(x') + \frac{1}{N} \sum_{n}\int a^{(v)}_{n} K_{\Delta_{n}}(x,z \vert x' ){\cal O}^{(v)}_{n}(x')
\end{equation}
This will make $\langle \phi_{1}(x,z)j_{\mu}(y_1){\cal O}_{2}(y_2)\rangle$ analytic for $\chi>0$, and thus produce a local field in the bulk, provided we set
\begin{equation}
a_{n}^{(v)}c_{n}^{(v)}=\alpha^{(v)} (a_{n}^{(s)}c_{n}^{(s)})
\end{equation}
Here $\alpha^{(v)}$ is a constant reflecting the different normalization of the 3-point we start with.

This bulk field obeys an equation of motion
\begin{equation}
(\nabla^2-\Delta_{1}(\Delta_{1}-d))\phi_{1}(x,z)= \frac{1}{N}\sum_{n}\int a^{(v)}_{n}[\Delta_{n}(\Delta_{n}-d)-\Delta_{1}(\Delta_{1}-d)] K_{\Delta_{n}}(x,z \vert x' ){\cal O}^{(v)}_{n}(x')
\label{eommassv}
\end{equation} 
To identify the right hand side we insert it in a 3-point function with $j_{\mu}(y_1)$ and ${\cal O}_{2}(y_2)$ and obtain
\begin{eqnarray}
& & \big((y_2-y_1)^2\partial^{y_1}_{\mu}-2\Delta(y_2-y_1)_{\mu}\big)\Big[\sum_{n}\int dx' a^{(v)}_{n}c^{(v)}_{n} \nonumber \\
&  &  [\Delta_{n}(\Delta_{n}-d)-\Delta_{1}(\Delta_{1}-d)] K_{\Delta_{n}}(x,z \vert x' )\langle \Delta_{n}, \Delta, \Delta_{2}+1\rangle \Big]
\end{eqnarray}
This is just $\alpha^{(v)} \left((y_2-y_1)^2\partial^{y_1}_{\mu}-2\Delta(y_2-y_1)_{\mu}\right)$ acting on the result of the scalar sum.
We have seen many examples of these scalar sums, so up to a normalization we know this will become
\begin{equation}
\left((y_2-y_1)^2\partial^{y_1}_{\mu}-2\Delta(y_2-y_1)_{\mu}\right)\left(\frac{z}{(x-y_1)^2+z^2}\right)^{\Delta}
\left(\frac{z}{(x-y_1)^2+z^2}\right)^{\Delta_{2}+1}
\label{massum}
\end{equation}
Using results from \cite{Kabat:2012hp}
\begin{eqnarray}
&&\langle zA_{\nu}(x,z)j_{\mu}(y_1)\rangle =\frac{\Delta-1}{\Delta}\eta_{\mu \nu} \left(\frac{z}{(x-y_1)^2+z^2}\right)^{\Delta}-\frac{z^\Delta}{2\Delta(\Delta-1)}\partial^{x}_{\mu}\partial^{x}_{\nu}\left(\frac{1}{(x-y_1)^2+z^2}\right)^{\Delta-1}\nonumber\\
&&\langle A_{z}(x,z)j_{\mu}(y_1)\rangle =\frac{1}{\Delta}\partial_{x_{\mu}} \left(\frac{z}{(x-y_1)^2+z^2}\right)^{\Delta}
\end{eqnarray}
and a little algebra (\ref{massum}) can be seen to be proportional to
\be
z^2\langle A^{(0)}_{\nu}(x,z)j_{\mu}(y_1)\rangle \eta^{\nu\rho}\partial_{x_{\rho}}\langle \phi^{(0)}_{2}(x,z){\cal O}_{2}(y_2)\rangle +
z^2\langle A^{(0)}_{z}(x,z)j_{\mu}(y_1)\rangle \partial_{z}\langle \phi^{(0)}_{2}(x,z){\cal O}_{2}(y_2)\rangle
\ee
This means the CFT equation of motion (\ref{eommassv}) for the bulk field is
\begin{equation}
(\nabla^2-\Delta_{1}(\Delta_{1}-d))\phi_{1}(x,z) \sim \frac{1}{N}A^{(0)}_{M}\partial^{M}\phi^{(0)}_{2}(x,z)
\end{equation}
which is indeed the expected bulk equation of motion to this order in $1/N$.
 
\section{Scalar coupled to gravity\label{sect:gravity}}
The three-point function of the energy-momentum tensor and two primary scalars of dimension $\Delta$  is given by
\begin{eqnarray}
\label{3pointt}
& &\langle {\cal O}(x){\cal O}(y_1)T_{\mu \nu}(y_2)\rangle = \frac{c_{d,\Delta}}{(y_2-y_1)^{d-2}(y_2-x)^{d-2}(x-y_1)^{2\Delta-d+2}}\\
& &\left[\left(\frac{(y_2-x)_{\mu}}{(y_2-x)^2}-\frac{(y_2-y_1)_{\mu}}{(y_2-y_1)^2}\right)\left(\frac{(y_2-x)_{\nu}}{(y_2-x)^2}-\frac{(y_2-y_1)_{\nu}}{(y_2-y_1)^2}\right)
-\frac{\eta_{\mu \nu}}{d}\left(\frac{(y_2-x)_{\rho}}{(y_2-x)^2}-\frac{(y_2-y_1)_{\rho}}{(y_2-y_1)^2}\right)^2\right]\nonumber
\end{eqnarray}
Defining a boundary Weyl tensor
\begin{equation}
C_{\alpha \beta \gamma \delta}=\partial_{\alpha} \partial_{\gamma}T_{\beta \delta}-(\gamma \leftrightarrow \delta)-(\beta \leftrightarrow \alpha)+(\gamma \leftrightarrow \delta \ \ \beta \leftrightarrow \alpha)
\end{equation}
and a differential operator
\begin{equation}
L_{\alpha \beta \gamma \delta}=(y_1-y_2)_{\alpha}(y_1-y_2)_{\gamma}\partial^{y_2}_{\beta}\partial^{y_2}_{\delta}-(\gamma \leftrightarrow \delta)-(\beta \leftrightarrow \alpha)+(\gamma \leftrightarrow \delta \ \ \beta \leftrightarrow \alpha)
\end{equation}
the three-point function of $C_{\alpha \beta \gamma \delta}$  where $\alpha, \beta, \gamma ,\delta$ are all distinct with two primary scalars of dimension $\Delta$ is\footnote{correcting a normalization factor in \cite{Kabat:2013wga} }
\begin{eqnarray}
 \langle {\cal O}(x){\cal O}(y_1)C_{\alpha \beta \gamma \delta}(y_2) \rangle =
\gamma_{c}^{g} L_{\alpha \beta \gamma \delta}\langle \Delta,\Delta+2,d\rangle 
\end{eqnarray}
where
\begin{equation}
\gamma_{c}^{g}=\frac{4(d-1)c_{d,\Delta}}{d-2}
\end{equation}
and the scalar correlator $\langle \Delta,\Delta+2,d\rangle$ is defined in (\ref{D1D2D3def}).
 $\gamma_{c}^{g}$ can be computed from the 3-point function of two scalars with an energy-momentum tensor.  The results in \cite{Liu:1998bu} imply
\begin{equation}
\gamma_{c}^{g}=-\kappa\frac{d(\Delta-\frac{d}{2})\Gamma(\frac{d}{2})\Gamma(\Delta+1)}{(d-2)\pi^{d}\Gamma(\Delta-\frac{d}{2})}
\end{equation}
where $\kappa = {\cal O}(1/N)$ is the gravitational coupling.

In \cite{Kabat:2013wga} a method for constructing a bulk scalar field $\phi$ coupled to gravity by smearing a dimension $\Delta$ primary operator ${\cal O}$ was
developed.  The method is based on enforcing an appropriate notion of bulk locality.  Due to the gravitational constraint equations one cannot require that all
commutators vanish at spacelike separation.  However in \cite{Kabat:2013wga} it was shown that the commutator of a bulk scalar with certain components of the
boundary Weyl tensor vanishes.  Thus the bulk field can be defined by
\begin{equation}
\phi(x,z)=\int dx' K_{\Delta}(x,z \vert x') {\cal O}(x')+\sum^{\infty}_{n=0} a^{(g)}_{n}\int dx' K_{\Delta_{n}}(x,z \vert x') {\cal T}_{\Delta_{n}}(x')
\label{bulkfg}
\end{equation}
where the scalar (but not primary) operators ${\cal T}_{n}$ are double-trace operators built from a product of $T_{\mu  \nu}$ and ${\cal O}$ with $2n+2$ derivatives.
They have dimension $\Delta_{n}=2n+\Delta+d+2$ and are constructed so that
\begin{equation}
\langle  {\cal T}_{n}(x) {\cal O}(y_1) C_{\alpha \beta \gamma \delta}(y_2)\rangle =c_{n}^{(g)}L_{\alpha \beta \gamma \delta}\langle \Delta_{n}, \Delta+2,d\rangle 
\label{3pt}
\end{equation}
where $\alpha, \beta, \gamma ,\delta$ are all distinct.\footnote{This means the construction only works for $d\geq 4$.}  To fix the coefficients $a^{(g)}_{n}$, we require
that the 3-point function of $\phi$ with ${\cal O}$ and $C_{\alpha \beta \gamma \delta}$ is analytic for $\chi >0$.

It's clear that all we have to do is to solve the related scalar system with operators of dimension $\Delta, \Delta+2,d$.
Having found the coefficients $a_{n}^{(s)}c_{n}^{(s)}$ that make the scalar system local in the bulk, all we need to do is set
\be
a_{n}^{(g)}c_{n}^{(g)}=\alpha^{g} a_{n}^{(s)}c_{n}^{(s)}
\ee
where
\be
\alpha^{g}=\frac{\gamma_{c}^{g}}{\gamma^{(d)}_{\Delta,\Delta+2,d}}
\label{coefsg}
\ee
takes into account the different normalization of the starting 3-point function.  The bulk field defined in (\ref{bulkfg}) will then be local with respect to
${\cal O}$ and $C_{\alpha\beta\gamma\delta}$.

The scalar system we need to solve is treated in appendix \ref{appendix:scalargravity} for $d=4$ and $d=5$.
Assuming we have solved the scalar system, let's look at the resulting bulk equation of motion.  From (\ref{bulkfg}) the bulk field satisfies
($m^2=\Delta(\Delta-d)$)
\begin{equation}
(\nabla^2-m^2)\phi(x,z)=\sum^{\infty}_{n=0}a^{(g)}_{n} \int dx' (M_{n}^{2}-m^2) K_{\Delta_{n}}(x,z \vert x') {\cal T}_{\Delta_{n}}(x')
\label{geom1}
\end{equation}
To identify the operator on the right hand side we put it in a 3-point function with ${\cal O}(y_1)$ and $C_{\alpha \beta \gamma \delta}(y_2)$.
Using (\ref{3pt}) and (\ref{coefsg}) we see that the right hand side of (\ref{geom1})  is just
\begin{equation}
\alpha^{(g)} L_{\alpha \beta \gamma \delta}\big[\hbox{\rm scalar sum}\big]
\end{equation}
The scalar sum we know.  For a bulk coupling constant $\frac{\lambda}{N}$ it is just
\begin{equation}
\frac{\lambda}{N}\frac{\Gamma(\Delta+2)\Gamma(d)}{\pi^{d}\Gamma(\Delta+2-\frac{d}{2})\Gamma(\frac{d}{2})}\left(\frac{z}{(x-y_1)^2+z^2}\right)^{\Delta+2}\left(\frac{z}{(x-y_1)^2+z^2}\right)^{d}
\label{geom2}
\end{equation}
So the right hand side of (\ref{geom1}) is an operator whose 3-point function with ${\cal O}(y_1)$ and $C_{\alpha \beta \gamma \delta}(y_2)$ is given by 
\begin{equation}
\alpha^{(g)}\frac{\lambda}{N}\frac{\Gamma(\Delta+2)\Gamma(d)}{\pi^{d}\Gamma(\Delta+2-\frac{d}{2})\Gamma(\frac{d}{2})} L_{\alpha \beta \gamma \delta}\left(\frac{z}{(x-y_1)^2+z^2}\right)^{\Delta+2}\left(\frac{z}{(x-y_1)^2+z^2}\right)^{d}
\label{eomcftg}
\end{equation}

We now show that the operator is just $\kappa g^{\mu \rho} g^{\nu\sigma} h^{(0)}_{\mu \nu}\partial_{\rho}\partial_{\sigma}\phi^{(0)}(x,z)$, where to this order in
$1/N$ $g^{\mu\rho}$ is  the inverse AdS metric and the superscript ${}^{(0)}$ indicates a free field constructed using the zeroth order smearing function
(for notation see (\ref{eom})). We need to collect a few results.  First, with $\gamma^{(d)}_{\Delta,\Delta+2,d}$ given by (\ref{gijk}) we have
\begin{equation}
\alpha^{g}=8(\Delta-\frac{d}{2})(\Delta-\frac{d}{2}+1)\frac{N\kappa}{\lambda}
\end{equation}
Then using\footnote{This is valid when $\alpha, \beta, \gamma, \delta$ are all distinct.  For details see appendix \ref{appendix:hC}.}
\begin{eqnarray}
& &\langle z^2h^{(0)}_{\mu \nu}(x,z) C_{\alpha \beta \gamma \delta}(y_2)\rangle =\frac{\Gamma(d)}{\pi^{\frac{d}{2}}\Gamma(\frac{d}{2})}\big[(\eta_{\mu \beta}\eta_{\nu \delta}+\eta_{\nu \beta}\eta_{\mu \delta})\partial^{y_2}_{\alpha}\partial^{y_2}_{\gamma} -(\gamma \leftrightarrow \delta)-\nonumber\\
& & \qquad (\beta \leftrightarrow \alpha)+(\gamma \leftrightarrow \delta \ \ \beta \leftrightarrow \alpha)\big]
\left(\frac{z}{(x-y_2)^2+z^2}\right)^{d}
\label{ttbulk}
\end{eqnarray}
and
\begin{eqnarray}
\langle \partial_{\rho}\partial_{\sigma}\phi^{(0)}(x,z){\cal O}(y_1)\rangle & = & \frac{4\Gamma(\Delta+2)}{\pi^{\frac{d}{2}}\Gamma(\Delta-\frac{d}{2})}(x-y_1)_{\rho}(x-y_1)_{\sigma}\left(\frac{z}{z^2+(x-y_1)^2}\right)^{\Delta+2} \nonumber \\
& & \quad + (\hbox{\rm terms proportional to $\eta_{\rho\sigma}$})
\end{eqnarray}
From this we get\footnote{Note that due to the antisymmetry properties of $L_{\alpha\beta\gamma\delta}$ we can replace $(x-y_1)_{\rho} \rightarrow (y_1-y_2)_{\rho}$}
\begin{eqnarray}
& &\kappa \langle  g^{\mu \rho} g^{\nu\sigma} h^{(0)}_{\mu \nu}\partial_{\rho}\partial_{\sigma}\phi^{(0)}(x,z){\cal O}(y_1)C_{\alpha\beta\gamma\delta}(y_2)\rangle \\
& &= \kappa \frac{8\Gamma(d)\Gamma(\Delta+2)}{\pi^{d}\Gamma(\frac{d}{2})\Gamma(\Delta-\frac{d}{2})}\left(\frac{z}{z^2+(x-y_1)^2}\right)^{\Delta+2}L_{\alpha\beta\gamma\delta}\left(\frac{z}{z^2+(x-y_2)^2}\right)^{d}\nonumber
\end{eqnarray}
which matches (\ref{eomcftg}). This means the bulk field satisfies an equation of motion which to this order in $1/N$ is
\begin{equation}
[\nabla^2-m^2]\phi(x,z)=\kappa g^{\mu \rho} g^{\nu\sigma} h^{(0)}_{\mu \nu}\partial_{\rho}\partial_{\sigma}\phi^{(0)}(x,z)
\end{equation}
This is the expected equation of motion to order $1/N$ in holographic gauge.

\bigskip
\bigskip
\goodbreak
\centerline{\bf Acknowledgements}
\noindent
We are grateful to Tom Banks for pointing out the importance of field redefinitions and to participants in the KITP focus week ``Decoding the hologram?'' for valuable feedback.  We thank the City College physics department for hospitality during this work.
DK is supported by U.S.\
National Science Foundation grant PHY-1125915.  The work of GL was supported in part by the Israel Science
Foundation under grant  504/13 and in part by a grant from
GIF, the German-Israeli Foundation for Scientific Research and
Development, under grant 1156-124.7/2011.

\appendix
\section{Bulk construction of local operators\label{appendix:bulk}}
In the bulk of this paper we used locality as a guiding principle.  We constructed local observables in the bulk starting from CFT correlators and we showed that these local
observables obeyed the expected bulk equations of motion.

Another way to construct local fields in the bulk is to solve the radial bulk evolution equations perturbatively \cite{Kabat:2011rz,Heemskerk:2012mn}.
In this appendix we adopt this alternate approach and show that it agrees with the approach based on locality, in the sense that for scalar fields it leads to the same
$a_n^{CFT}$ coefficients appearing in (\ref{coefbulk}).

Suppose we have a bulk equation of motion for a scalar field in AdS${}_{d+1}$ with a cubic coupling of the form\footnote{Other forms of cubic bulk interaction are discussed in section \ref{sect:redef}.}
\begin{equation}
\nabla^2 \phi_{i} -m^2 \phi_{i} =\frac{\lambda}{N} \phi_{j}\phi_{k}
\end{equation}
Here $\nabla^2=\frac{1}{\sqrt{g} }\partial_{M} \sqrt{g} g^{MN} \partial_{N}$ and $\lambda$ is a coupling which is ${\cal O}(1)$.
We can solve this in perturbation theory by setting $\phi =\phi^{(0)} +\frac{1}{N}\phi^{(1)} + \cdots$
and gathering powers of $1/N$.  At ${\cal O}(N^0)$ we find that $\phi^{(0)}$ is a free field,
\be
(\nabla^2 -m^2)\phi_{i}^{(0)}=0\,,
\label{eom1}
\ee
while at ${\cal O}(1/N)$ we find the correction
\be
(\nabla^2 -m^2)\phi_{i}^{(1)}=\lambda \phi_{j}^{(0)}\phi_{k}^{(0)}\,.
\label{eom2}
\ee
We want to solve these equations in terms of data given on the time-like boundary of AdS, namely that
the bulk field $\phi$ should behave near the boundary as
\begin{equation}
\phi_{i}(x,z) \sim z^{\Delta} {\cal O}_{i}(x) \qquad \hbox{\rm as $z \rightarrow 0$}
\label{z0lim}
\end{equation}
Here $m^2=\Delta(\Delta-d)$ and ${\cal O}_i(x)$ is identified with an operator in the CFT.

This is not a standard Cauchy problem but nevertheless it can be solved.
The solution of the free wave equation (\ref{eom1}) is
\begin{equation}
\phi^{(0)}_{i}(x,z)=\int d^dx' K_{\Delta_{i}}(x,z \vert x') {\cal O}_{i}(x')
\label{eom}
\end{equation}
where the smearing kernel  $K_{\Delta}(x,z \vert x')$ obeys
\begin{equation}
\nabla^2 K_{\Delta}(x,z \vert x')=\Delta(\Delta-d) K_{\Delta}(x,z \vert x')\,.
\end{equation}
The second equation (\ref{eom2}) can be solved with a help of a Green's function obeying
\begin{equation}
(\nabla^2 - m^2)G(X,X')=\frac{1}{\sqrt{g}}\delta(X-X')
\label{greeneq}
\end{equation}
to give
\begin{equation}
\phi^{(1)}_{i}(X)=\lambda \int d^{d+1}X' \sqrt{g} \, G(X,X') \, \phi_{j}^{(0)} \phi_{k}^{(0)}(X')
\label{greenbulk}
\end{equation}
where $X$ and $X'$ label bulk points.

There is an alternate way of solving this equation which is easier to compare to CFT results.
Note that since $\phi_{j}^{(0)}\phi_{k}^{(0)}(X)$ is a local scalar in the bulk it must be possible to write it as a  sum of smeared primary double-trace operators ${\cal O}_n$ in the CFT.
(The ${\cal O}_n$'s are built from a product of ${\cal O}_{j}$ and ${\cal O}_{k}$ with $2n$ derivatives and have dimension $\Delta_{n}=\Delta_j + \Delta_{k} +2n$.) So we can write
\begin{equation}
\phi_{j}^{(0)}\phi_{k}^{(0)}(X)=\sum_{n} b^{jk}_n \int d^dx' K_{\Delta_n}(x,z \vert x'){\cal O}_{n}(x')
\label{phi0phi0}
\end{equation}
Explicit formulas for $b^{jk}_n$ have been determined using the K\"all\'en-Lehmann representation for AdS by Bros et al.\ \cite{Bros:2011vh}.

In this spirit we write 
\begin{equation}
\phi^{(1)}_{i}(X)=\sum_{n} a^{bulk}_{n} \int d^dx' K_{\Delta_n}(x,z \vert x'){\cal O}_{n}(x')
\label{phi1}
\end{equation}
and insert (\ref{phi1}) and (\ref{phi0phi0}) in (\ref{eom2}).  Labeling $M_{n}^2=\Delta_{n}(\Delta_{n}-d)$  we get
\begin{equation}
\sum_{n} a^{bulk}_{n} (M_{n}^2-m^2)\int d^dx' K_{\Delta_n}(x,z \vert x'){\cal O}_{n}(x')=\lambda \sum_{n} b^{jk}_n \int d^dx' K_{\Delta_n}(x,z \vert x'){\cal O}_{n}(x')
\end{equation}
Since the ${\cal O}_n$ are orthogonal inside a two-point function we can read off
\begin{equation}
\label{anexpression}
a^{bulk}_{n} = \frac{\lambda \, b^{jk}_{n}}{M_{n}^2-m^2}\,.
\end{equation}
These coefficients determine the interacting bulk field to ${\cal O}(1/N)$.

The solution we have obtained is of course consistent with the one obtained from a Green's function in (\ref{greenbulk}).
A useful identity is
\begin{equation}
\label{UsefulIdentity}
\int d^{d+1} X' \sqrt{g} G(X,X') \int d^dx'' K_{\Delta_{n}}(X' \vert x''){\cal O}_{n}(x'') = \frac{1}{M_{n}^2-m^2} \int d^dx'' K_{\Delta_{n}}(X \vert x''){\cal O}_{n}(x'')
\end{equation}
(To prove this act on both sides with $\nabla^2 - m^2$ and use the fact that the Green's function obeys
(\ref{greeneq}) while the smearing kernel obeys the free equation of motion.)  Substituting (\ref{phi0phi0}) in (\ref{greenbulk}) and using (\ref{UsefulIdentity}) reproduces (\ref{anexpression}).

\subsection{Computing $a_n c_{njk}$}
The coefficients $b^{jk}_{n}$ have been computed \cite{Bros:2011vh}, and in principle from (\ref{anexpression}) this result determines $a_n^{bulk}$.
But when comparing bulk results to the CFT,
it is easier to compare the combination $a_n c_{njk}$ where $c_{njk}$ is the coefficient appearing in the 3-point function
$\langle {\cal O}_n {\cal O}_j {\cal O}_k \rangle$.

We'll first determine $b^{jk}_{n}c_{njk}$ then determine $a^{bulk}_{n}c_{njk}$.  The combination $b^{jk}_{n}c_{njk}$ can be computed in the following way.
Consider the bulk-boundary 3-point function
\begin{equation}
\langle \phi^{(0)}_{j}\phi^{(0)}_{k}(x,z){\cal O}_{j}(y_1){\cal O}_{k}(y_2)\rangle _{CFT}
\label{coef1}
\end{equation}
where 
\begin{eqnarray}
\phi_{j}^{(0)}(x,z)=\int dx' K_{\Delta_j}(x,z \vert x'){\cal O}_{j}(x')\\
\phi_{k}^{(0)}(x,z)=\int dx' K_{\Delta_k}(x,z \vert x'){\cal O}_{k}(x')\nonumber
\end{eqnarray}
This can be evaluated by two means at leading order for large $N$.  One can use large $N$ factorization to get
\begin{eqnarray}
& &\langle \phi^{(0)}_{j}(x,z){\cal O}_{j}(y_1)\rangle \langle \phi^{(0)}_{k}(x,z){\cal O}_{k}(y_2)\rangle =\nonumber\\
& &d_{jk}\left(\frac{z}{(x-y_1)^2+z^2}\right)^{\Delta_j}\left(\frac{z}{(x-y_2)^2+z^2}\right)^{\Delta_k}
\label{coef2}
\end{eqnarray}
where
\begin{equation}
d_{jk}=\frac{1}{\pi^{d}}\frac{\Gamma(\Delta_{j})\Gamma(\Delta_{k})}{\Gamma(\Delta_{j}-\frac{d}{2})\Gamma(\Delta_{k}-\frac{d}{2})}
\label{djk}
\end{equation}
Alternatively we can use the fact that $\phi^{(0)}_{j}\phi^{(0)}_{k}(x,z)$ is a local bulk scalar to write
\begin{equation}
\phi^{(0)}_{j}\phi^{(0)}_{k}(x,z)=\sum_{n} b^{jk}_{n} \int dx' K_{\Delta_n}(x,z \vert x'){\cal O}_{n}(x')
\end{equation}
where ${\cal O}_{n}$ is a primary scalar double-trace operator built from ${\cal O}_{j}$ and ${\cal O}_{k}$ with $2n$ derivatives.  It has dimension
$\Delta_{n}=\Delta_{j}+\Delta_{k}+2n$.  So using the result (\ref{3pointbulk}), (\ref{coef1}) can also be written as 
\begin{eqnarray}
& &\sum_{n} b^{jk}_{n}c_{njk}\frac{1}{(y_1 -y_2)^{2\Delta_{j}}}
\left[\frac{z}{z^2+(x-y_2)^2}\right]^{(\Delta_{k}-\Delta_{j})} \nonumber\\
& &\times \left(\frac{1}{\chi-1}\right)^\DS F(\DS,\DS-\frac{d}{2}+1,\Delta_{n}-\frac{d}{2}+1,\frac{1}{1-\chi})
\label{coef3}
\end{eqnarray}
Equating  expressions (\ref{coef2}) and (\ref{coef3}) 
\begin{eqnarray}
\label{coef3b}
& &d_{jk}\left(\frac{z}{(x-y_1)^2+z^2}\right)^{\Delta_j}\left(\frac{z}{(x-y_2)^2+z^2}\right)^{\Delta_k} \\
& & =\sum_{n} b^{jk}_{n}c_{njk}\frac{1}{(y_1 -y_2)^{2\Delta_{j}}}
\left[\frac{z}{z^2+(x-y_2)^2}\right]^{(\Delta_{k}-\Delta_{j})} \nonumber\\
\nonumber
& &\times \left(\frac{1}{\chi-1}\right)^\DS F(\DS,\DS-\frac{d}{2}+1,\Delta_{n}-\frac{d}{2}+1,\frac{1}{1-\chi})
\end{eqnarray}
one can read off the coefficients $b^{jk}_{n}c_{njk}$.
In practice this is easiest to do in the limit $z\rightarrow 0$, but then we can only easily read off the coefficient with the smallest value of $n$ on the right hand side. To
determine the coefficients for larger $n$ we act on both sides of the equation with $\nabla^{2}-\Delta_{m}(\Delta_{m}-d)$, which for appropriate choices of $\Delta_{m}=2m+\Delta_{j}+\Delta_{k}$ will eliminate one after another of the leading terms on the right hand side. So to easily read off the coefficient for a particular value of
$n$ we act on both sides with the operator
\begin{equation}
\prod_{m=0}^{n-1}\left(\nabla^{2}-\Delta_{m}(\Delta_{m}-d)\right)
\end{equation}
Let's start with $n = 1$ and see what this operator does to (\ref{coef3b}). On the left hand side, using
\begin{equation}
\nabla^2=z^2\partial_\mu \partial^\mu+(1-d)z\partial_{z}+z^2\partial^{2}_{z}
\end{equation}
and since for any $\Delta$,
\begin{equation}
(\nabla^2 -\Delta(\Delta-d))\left(\frac{z}{(x-y_1)^2+z^2}\right)^{\Delta}=0
\end{equation}
we get
\begin{eqnarray}
& &(\nabla^2 -(\Delta_j +\Delta_k)((\Delta_j+\Delta_{k})-d))\left(\frac{z}{(x-y_1)^2+z^2}\right)^{\Delta_j}\left(\frac{z}{(x-y_1)^2+z^2}\right)^{\Delta_k}\nonumber\\
& & = -4\Delta_{j}\Delta_{k}(y_1-y_2)^2\left(\frac{z}{(x-y_1)^2+z^2}\right)^{\Delta_j +1}\left(\frac{z}{(x-y_1)^2+z^2}\right)^{\Delta_k+1}
\end{eqnarray}
On the right hand side of (\ref{coef3b}) this operator eliminates the $n=0$ term and multiplies all other terms by $\Delta_{n}(\Delta_{n}-d)-\Delta_{0}(\Delta_{0}-d)$. Then taking the limit $z \rightarrow 0$ we can easily read off the $b^{jk}_{1}c_{1jk}$ coefficient. This procedure can be repeated again and again, and we get
\begin{equation}
\prod_{m=0}^{n-1}[\Delta_{n}(\Delta_{n}-d)-\Delta_{m}(\Delta_{m}-d)] b^{jk}_{n}c_{njk}=d_{jk}(-4)^{n}\frac{\Gamma(\Delta_{j}+n)\Gamma(\Delta_{k}+n)}{\Gamma(\Delta_{j})\Gamma(\Delta_{k})}
\end{equation}
Using
\begin{equation}
\Delta_{n}(\Delta_{n}-d)-\Delta_{m}(\Delta_{m}-d)=4(n-m)(n+m+\Delta_{j}+\Delta_{k}-\frac{d}{2})
\end{equation}
this can be written in the convenient form
\begin{equation}
b^{jk}_{n}c_{njk}=\frac{1}{\pi^{d}}\frac{(-1)^{n}}{\Gamma(\Delta_{j}-\frac{d}{2})\Gamma(\Delta_{k}-\frac{d}{2})} \frac{\Gamma(\Delta_{j}+n)\Gamma(\Delta_{k}+n)\Gamma(n+\Delta_{j}+\Delta_{k}-\frac{d}{2})}{\Gamma(n+1)\Gamma(2n+\Delta_{j}+\Delta_{k}-\frac{d}{2})}
\end{equation}

Thus for the case of a bulk $\frac{\lambda}{N}\phi_{i}\phi_{j}\phi_{k}$ interaction, solving the bulk equations of motion gives a first-order
correction to the field which can be put in the form (\ref{phi1}).  From (\ref{anexpression}), the coefficients $a^{bulk}_{n}$ obey
\begin{eqnarray}
a^{bulk}_{n}c_{njk}&=&\frac{\lambda}{\pi^{d}}\frac{(-1)^{n}}{\Gamma(\Delta_{j}-\frac{d}{2})\Gamma(\Delta_{k}-\frac{d}{2})}
\frac{\Gamma(\Delta_{j}+n)\Gamma(\Delta_{k}+n)\Gamma(n+\Delta_{j}+\Delta_{k}-\frac{d}{2})}{\Gamma(n+1)\Gamma(2n+\Delta_{j}+\Delta_{k}-\frac{d}{2})}\nonumber\\
& & \times \frac{1}{\Delta_{n}(\Delta_{n}-d)-\Delta_{i}(\Delta_{i}-d)}\nonumber
\label{coefbulk2}
\end{eqnarray}
which agrees with the general form (\ref{coefbulk}).
This result was obtained from the bulk point of view and relied on knowing the bulk equations of motion.  We have shown in numerous examples that the same result can be obtained purely within the CFT
by the requirement of bulk locality.

\section{Poles and cuts in even dimensions\label{appendix:polecut}}
In the section we treat the case of even $d \geq 4$, and show that cancellation of the unwanted cuts in $I_{ijk}$ automatically leads to cancellation of the
unwanted poles.

We start with a 3-point function of scalar primaries of dimension $\Delta_{\nu},\Delta_{j},\Delta_k$. We want to promote the operator of dimension $\Delta_{\nu}$ to a bulk field, using higher-dimension double-trace primaries that have dimension $\Delta_{n}=2n+\Delta_{j}+\Delta_{k}$ to achieve locality.  Define
\begin{eqnarray}
\Delta_{0}^{(\nu)}&=&\frac{1}{2}(\Delta_{\nu}+\Delta_{j}-\Delta_k)\nonumber\\
\Delta_{0}^{(n)}&=&\frac{1}{2}(\Delta_{n}+\Delta_{j}-\Delta_k)
\end{eqnarray}
The discontinuity across the cut for the original 3-point function is given by (\ref{discont}) and (\ref{Evendgeq4Discont}).
\begin{equation}
\tilde{\gamma}\frac{\Gamma(\Delta_{\nu}-1)}{\Gamma(\Delta_{0}^{(\nu)})\Gamma(\Delta_{\nu}-\Delta_{0}^{(\nu)})}(\Delta_{0}^{(\nu)}-1)(1+\Delta_{0}^{(\nu)}-\Delta_{\nu})F(\Delta_{0}^{(\nu)},2+\Delta_{0}^{(\nu)}-\Delta_{\nu},2,1-\chi)
\label{orcutd5}
\end{equation}
In even $d \geq 4$ there are also poles in the 3-point function at $\chi = 1$, coming from
\begin{equation}
F(a+m,a,c,z) = (-z)^{-a}\frac{\Gamma(c)}{\Gamma(a+m)}\sum_{l=0}^{m-1}\frac{\Gamma(m-l)(a)_{l}}{\Gamma(c-a-l)}z^{-l} + \cdots
\label{polepart}
\end{equation}
where $(a)_{l}=\frac{\Gamma(a+l)}{\Gamma(a)}$. 

We start with the case $d = 4$.  Applying (\ref{polepart}) to (\ref{Idef}), in $d = 4$ the pole is given by
\begin{equation}
\tilde{\gamma}\frac{\Gamma(\Delta_{\nu}-1)}{\Gamma(\Delta_{0}^{(\nu)})\Gamma(\Delta_{\nu}-\Delta_{0}^{(\nu)})}\frac{1}{\chi-1}
\end{equation}
When an operator of dimension $\Delta_n$ is inserted in the correlator in place of the operator with dimension $\Delta_\nu$, in place of (\ref{orcutd5})
the discontinuity across the cut is given by
\begin{equation}
c_{njk}\frac{\Gamma(\Delta_{n}-1)}{\Gamma(\Delta_{0}^{(n)})\Gamma(\Delta_{n}-\Delta_{0}^{(n)})}(\Delta_{0}^{(n)}-1)(1+\Delta_{0}^{(n)}-\Delta_{n})F(\Delta_{0}^{(n)},2+\Delta_{0}^{(n)}-\Delta_{n},2,1-\chi)
\label{highcutd5}
\end{equation}
while the pole is given by
\begin{equation}
c_{njk}\frac{\Gamma(\Delta_{n}-1)}{\Gamma(\Delta_{0}^{(n)})\Gamma(\Delta_{n}-\Delta_{0}^{(n)})}\frac{1}{\chi-1}
\end{equation}

Now imagine we have found coefficients $a_{n}$ such that
\begin{eqnarray}
\label{d5cancel}
& & \tilde{\gamma}\frac{\Gamma(\Delta_{\nu}-1)}{\Gamma(\Delta_{0}^{(\nu)})\Gamma(\Delta_{\nu}-\Delta_{0}^{(\nu)})}F(\Delta_{0}^{(\nu)},2+\Delta_{0}^{(\nu)}-\Delta_{\nu},2,1-\chi)\\
& &=-\frac{1}{N}\sum_{n=0}^{\infty} a_{n}c_{njk}\frac{\Gamma(\Delta_{n}-1)}{\Gamma(\Delta_{0}^{(n)})\Gamma(\Delta_{n}-\Delta_{0}^{(n)})}F(\Delta_{0}^{(n)},2+\Delta_{0}^{(n)}-\Delta_{n},2,1-\chi)\nonumber
\end{eqnarray}
Evaluating this at $\chi=1$ we see that the poles associated with higher dimension operators will exactly cancel the original pole.
To see that the same coefficients will also cancel the discontinuity across the cut we note
that the hypergeometric function obeys the differential equation
\begin{equation}
\left(z(1-z)\frac{d^2}{dz^2}+[c-(a+b+1)z]\frac{d}{dz}-ab\right)F(a,b,c,z)=0.
\end{equation}
Define the operator
\begin{equation}
L_{F,d}^{(p)}=L_{F,d}+(p-\frac{d}{2})(p+\Delta_{j}-\Delta_{k})
\end{equation}
where $L_{F,d}$ is the differential operator ($z=1-\chi$)
\begin{equation}
z(1-z)\frac{d^2}{dz^2}+[\frac{d}{2}-(1+\frac{d}{2}+\Delta_{j}-\Delta_{k})z]\frac{d}{dz}\,.
\end{equation}
Acting on (\ref{d5cancel}) with $L^{(1)}_{F,d}$ and using
\begin{eqnarray}
L^{(1)}_{F,4}F(\Delta_{0}^{(\nu)},2+\Delta_{0}^{(\nu)}-\Delta_{\nu},2,z)&=&(\Delta_{0}^{(\nu)}-1)(1+\Delta_{0}^{(\nu)}-\Delta_{\nu})F(\Delta_{0}^{(\nu)},2+\Delta_{0}^{(\nu)}-\Delta_{\nu},2,z)\nonumber\\
L^{(1)}_{F,4}F(\Delta_{0}^{(n)},2+\Delta_{0}^{(n)}-\Delta_{n},2,z)&=&
(\Delta_{0}^{(n)}-1)(1+\Delta_{0}^{(n)}-\Delta_{n})F(\Delta_{0}^{(n)},2+\Delta_{0}^{(n)}-\Delta_{n},2,z)\nonumber
\end{eqnarray}
we see by looking at (\ref{orcutd5}) and (\ref{highcutd5}) that the coefficients in
(\ref{d5cancel}) will also cancel the discontinuity across the cut.
Thus we have shown that cancellation of the discontinuity across the cut (which is easier to actually compute) implies the cancellation of the unwanted poles.

The case for even  $d>4$ is similar.
The discontinuity across the cut for the starting 3-point function is given by (\ref{discont}) and (\ref{Evendgeq4Discont})
\begin{equation}
\frac{\tilde{\gamma}}{(\frac{d}{2}-1)!}\frac{\Gamma(\Delta_{\nu}-\frac{d}{2}+1)}{\Gamma(\Delta_{0}^{(\nu)})\Gamma(\Delta_{\nu}-\Delta_{0}^{(\nu)})}(\Delta_{0}^{(\nu)}-\frac{d}{2}+1)_{\frac{d}{2}-1}(1+\Delta_{0}^{(\nu)}-\Delta_{\nu})_{\frac{d}{2}-1}F(\Delta_{0}^{(\nu)},\frac{d}{2}+\Delta_{0}^{(\nu)}-\Delta_{\nu},\frac{d}{2},1-\chi)
\label{orcutdg}
\end{equation}
while the poles (\ref{polepart}) can be written as
\begin{equation}
\frac{\tilde{\gamma}}{(\chi-1)^{\frac{d}{2}-1}}\frac{\Gamma(\Delta_{\nu}-\frac{d}{2}+1)}{\Gamma(\Delta_{0}^{(\nu)})\Gamma(\Delta_{\nu}-\Delta_{0}^{(\nu)})}\sum_{l=0}^{\frac{d}{2}-2}\frac{\Gamma(\frac{d}{2}-1-l)(\Delta_{0}^{(\nu)}-\frac{d}{2}+1)_{l}(\Delta_{0}^{(\nu)}-\Delta_{\nu}+1)_{l}}{l!}(\chi-1)^l
\label{orgpoldg}
\end{equation}
The contribution of the higher-dimension double-trace operators to the cut is given by (\ref{discont}) and (\ref{Evendgeq4Discont}),
\begin{equation}
\frac{c_{njk}}{(\frac{d}{2}-1)!}\frac{\Gamma(\Delta_{n}-\frac{d}{2}+1)}{\Gamma(\Delta_{0}^{(n)})\Gamma(\Delta_{n}-\Delta_{0}^{(n)})}(\Delta_{0}^{(n)}-\frac{d}{2}+1)_{\frac{d}{2}-1}(1+\Delta_{0}^{(n)}-\Delta_{n})_{\frac{d}{2}-1}F(\Delta_{0}^{(n)},\frac{d}{2}+\Delta_{0}^{(n)}-\Delta_{n},\frac{d}{2},1-\chi)
\label{highcutdg}
\end{equation}
while their contribution to the poles is given by (\ref{polepart}).
\begin{equation}
\frac{\tilde{\gamma}}{(\chi-1)^{\frac{d}{2}-1}}\frac{\Gamma(\Delta_{n}-\frac{d}{2}+1)}{\Gamma(\Delta_{0}^{(n)})\Gamma(\Delta_{n}-\Delta_{0}^{(n)})}\sum_{l=0}^{\frac{d}{2}-2}\frac{\Gamma(\frac{d}{2}-1-l)(\Delta_{0}^{(n)}-\frac{d}{2}+1)_{l}(\Delta_{0}^{(n)}-\Delta_{n}+1)_{l}}{l!}(\chi-1)^l
\label{highpoldg}
\end{equation}
Suppose we start with coefficients $a_{n}$ satisfying
\begin{eqnarray}
\label{basicevd}
&&\tilde{\gamma}\frac{\Gamma(\Delta_{\nu}-\frac{d}{2}+1)}{\Gamma(\Delta_{0}^{(\nu)})\Gamma(\Delta_{\nu}-\Delta_{0}^{(\nu)})}F(\Delta_{0}^{(\nu)},\frac{d}{2}+\Delta_{0}^{(\nu)}-\Delta_{\nu},\frac{d}{2},1-\chi)\\
&&=-\frac{1}{N}\sum_{n=0}^{\infty}a_{n}c_{njk}\frac{\Gamma(\Delta_{n}-\frac{d}{2}+1)}{\Gamma(\Delta_{0}^{(n)})\Gamma(\Delta_{n}-\Delta_{0}^{(n)})}F(\Delta_{0}^{(n)},\frac{d}{2}+\Delta_{0}^{(n)}-\Delta_{n},\frac{d}{2},1-\chi)
\nonumber
\end{eqnarray}
Evaluating this at $\chi=1$, we see that the $l = 0$ poles will cancel between the original 3-point function and the one involving double-trace operators.
To exhibit additional cancellations, note that
\begin{eqnarray}
&&(\Delta_{0}^{(\nu)}-\frac{d}{2}+p)(\Delta_{0}^{(\nu)}-\Delta_{\nu}+p)=\Delta_{0}^{(\nu)}(\Delta_{0}^{(\nu)}-\Delta_{\nu}+\frac{d}{2})+(p-\frac{d}{2})(\Delta_{j}-\Delta_{k}+p)\nonumber\\
&&(\Delta_{0}^{(n)}-\frac{d}{2}+p)(\Delta_{0}^{(n)}-\Delta_{n}+p)=\Delta_{0}^{(n)}(\Delta_{0}^{(n)}-\Delta_{n}+\frac{d}{2})+(p-\frac{d}{2})(\Delta_{j}-\Delta_{k}+p) \nonumber
\end{eqnarray}
This means
\begin{eqnarray}
& &L_{F,d}^{(p)}F(\Delta_{0}^{(\nu)},\frac{d}{2}+\Delta_{0}^{(\nu)}-\Delta_{\nu},\frac{d}{2},1-\chi)\\
& &= (\Delta_{0}^{(\nu)}-\frac{d}{2}+p)(\Delta_{0}^{(\nu)}-\Delta_{\nu}+p)F(\Delta_{0}^{(\nu)},\frac{d}{2}+\Delta_{0}^{(\nu)}-\Delta_{\nu},\frac{d}{2},1-\chi)
\nonumber
\end{eqnarray}
and
\begin{eqnarray}
& &L_{F,d}^{(p)}F(\Delta_{0}^{(n)},\frac{d}{2}+\Delta_{0}^{(n)}-\Delta_{n},\frac{d}{2},1-\chi)\\
& &=(\Delta_{0}^{(n)}-\frac{d}{2}+p)(\Delta_{0}^{(n)}-\Delta_{n}+p)F(\Delta_{0}^{(n)},\frac{d}{2}+\Delta_{0}^{(n)}-\Delta_{n},\frac{d}{2},1-\chi)
\nonumber
\end{eqnarray}
We see that if we act on (\ref{basicevd}) with $L_{F,d}^{(1)}$ we get the correct formula for canceling the $l = 1$ poles. Likewise acting with $L_{F,d}^{(1)}\cdots L_{F,d}^{(p)}$ we see that the $l = p$ poles cancel. Finally acting with $L_{F,d}^{(1)}\cdots L_{F,d}^{(\frac{d}{2}-1)}$ on (\ref{basicevd}), we see that the same
coefficients $a_{n}$ also cancel the cut.

\section{Scalar system relevant for gauge fields\label{appendix:scalargauge}}
The  starting 3-point function can be written as some derivative operator acting on a scalar 3-point function whose dimensions are (in $d=3$)
$\Delta, \Delta+1, 2$ (we will assume integer $\Delta$).
The discontinuity is given by (\ref{discont})
\begin{equation}
\tilde{\gamma}\frac{1}{\pi}\frac{1}{(1-\chi)^{1/2}}\left(\frac{1}{\chi}\right)^{\Delta-1}
\end{equation}
We can now build double-trace operators of dimension $\Delta_{n}=\Delta+3+2n$, which give a discontinuity
\begin{equation}
\frac{c_{n}}{\sqrt{\pi}}\frac{1}{(1-\chi)^{1/2}}\frac{\Gamma(\Delta+2n+2\frac{1}{2})}{\Gamma(n+\Delta+1)\Gamma(n+2)}F(\Delta+n+\frac{1}{2},-n-1,\frac{1}{2},1-\chi)
\end{equation}
This can be written using associated Legendre function as
\begin{equation}
\frac{c_{n}}{(1-\chi)^{1/2}}\frac{(-1)^{n+\Delta}\Gamma(\Delta+2n+2\frac{1}{2})}{2^{\Delta-1}\Gamma(n+\Delta+1)\Gamma(n+\Delta+\frac{1}{2})}\left(\frac{1}{\chi}\right)^{\frac{\Delta-1}{2}}P^{\Delta-1}_{2n+\Delta+1}(\sqrt{1-\chi})
\end{equation}
We see that in order to cancel the unwanted discontinuity, we need to be able to write
\begin{equation}
\left(\frac{1}{1-x^2}\right)^{\frac{\Delta-1}{2}}=\sum_{k}\alpha_{k}P_{k}^{\Delta-1}(x)
\end{equation}
To find $\alpha_{k}$ we need the integral
\begin{equation}
\int_{1}^{1}\left(\frac{1}{1-x^2}\right)^{\frac{m}{2}}P^{m}_{k}(x)
\end{equation}
This can be done by noticing that $P_{k}^{m}=(-1)^{m}(1-x^2)^{\frac{m}{2}}\frac{d^m}{dx^{m}}P_{k}(x)$,
so the integral is just
\begin{equation}
(-1)^m\frac{d^{m-1}}{dx^{m-1}}P_{k}(x)|_{-1}^{1}
\end{equation}
Using
\begin{equation}
P_{\nu}(x)=F(\nu,\nu+1,1,\frac{1-x}{2})
\end{equation}
we get 
\begin{equation}
\frac{d^{m-1}}{dx^{m-1}}P_{k}(1)=\frac{\Gamma(k+m)}{2^{m-1}(\Gamma(m)\Gamma(k-m+2)} \quad \hbox{\rm for $ k>m-2$}
\end{equation}
and zero otherwise.
So if $k$ and $m$ are both even or both odd we get
\begin{equation}
\int_{1}^{1}\left(\frac{1}{1-x^2}\right)^{\frac{m}{2}}P^{m}_{k}(x)=\frac{(-1)^{m}\Gamma(k+m)}{2^{m-2}\Gamma(m)\Gamma(k-m+2)} \quad \hbox{\rm for $ k>m-2$}
\end{equation}
Combined with
\begin{equation}
\int_{-1}^{1}P_{n}^{m}(x)P_{k}^{m}(x)=\delta_{nk}\frac{2\Gamma(n+m+1)}{(2n+1)\Gamma(n-m+1)}
\end{equation}
we get
\begin{equation}
\alpha_{k}=\frac{(-1)^{m}}{2^{m-1}\Gamma(m)}\frac{2k+1}{(k+m)(k-m+1)} \quad \hbox{\rm for $ k>m-2$}
\end{equation}
Note that putting $k=2n+\Delta+1$ and $m=\Delta-1$, we get
\begin{equation}
(k+m)(k-m+1)=(2n+2\Delta)(2n+3)=\Delta_{n}(\Delta_{n}-3)-\Delta(\Delta-3)
\end{equation}
So we can write
\begin{equation}
\frac{1}{\pi}\left(\frac{1}{\chi}\right)^{\frac{\Delta-1}{2}}=\frac{(-1)^{\Delta-1}}{\pi2^{\Delta-2}\Gamma(\Delta-1)}\sum_{n=-1}^{\infty}\frac{4n+2\Delta+3}{\Delta_{n}(\Delta_{n}-3)-\Delta(\Delta-3)}P^{\Delta-1}_{2n+\Delta+1}(\sqrt{1-\chi})
\label{basicjnotk}
\end{equation}
However we do not have the $n=-1$ term available, so we need to act with an operator that annihilates $P^{\Delta-1}_{\Delta-1}$. This is 
\begin{equation}
L_{\Delta-1}^{\Delta-1}=L_{\Delta-1}+\frac{(\Delta-1)^2}{1-x^2}
\end{equation}
Now
\begin{eqnarray}
& &L_{\Delta-1}^{\Delta-1}\left(\frac{1}{1-x^2}\right)^{\frac{\Delta-1}{2}}=(2-2\Delta)\left(\frac{1}{1-x^2}\right)^{\frac{\Delta-1}{2}}\\
& &L_{\Delta-1}^{\Delta-1}P_{2n+\Delta+1}^{\Delta-1}(x)=[(2n+\Delta+1)(2n+\Delta+2)-\Delta(\Delta-1)]P_{2n+\Delta+1}^{\Delta-1}(x)\nonumber\\
& = & 4(n+1)(n+\Delta+\frac{1}{2})P_{2n+\Delta+1}^{\Delta-1}(x)\nonumber
\end{eqnarray}
Putting this all together we have
\begin{equation}
a_{n}^{CFT}c_{n}=-N\tilde{\gamma}\frac{2^{3}}{\pi\Gamma(\Delta)}\frac{(-1)^{n}(n+1)\Gamma(n+\Delta+1)\Gamma(n+\Delta+\frac{3}{2})}{[\Delta_{n}(\Delta_{n}-3)-\Delta(\Delta-3)]\Gamma(2n+\Delta+\frac{3}{2})}
\end{equation}
where $\tilde{\gamma}$ is given by (\ref{corgijk})
\begin{equation}
\tilde{\gamma}=-\frac{\lambda}{8N\pi^2}\frac{\Gamma(\Delta)}{\Gamma(\Delta-\frac{1}{2})\Gamma(\frac{1}{2})}
\end{equation}
Again we find agreement with the general result (\ref{coefbulk}).

\subsection{Equations of motion}
Let us now sum the analytic part of the 3-point function to derive the equations of motion. From (\ref{3pointbulk}), (\ref{hypertrans1}), (\ref{anapartd}) the analytic part for the case $(\Delta,\Delta+1,2)$ has the form
\begin{equation}
c_{n}\frac{1}{(y_1-y_2)^{2\Delta+2}}\left[\frac{z}{z^2+(x-y_2)^{2}}\right]^{1-\Delta}\frac{\Gamma(-\frac{1}{2})\Gamma(2n+\Delta+2\frac{1}{2})}{\Gamma(n+\frac{3}{2})\Gamma(n+\Delta+\frac{1}{2})}F(n+\Delta+1, -n-\frac{1}{2},\frac{3}{2}, 1-\chi)
\end{equation}
We will use that for $-1<x<1$
\begin{equation}
F(n+\Delta+1, -n-\frac{1}{2},\frac{3}{2}, x^2)=\frac{(-1)^{n+\Delta}\Gamma(n+\frac{3}{2})}{2^{\Delta-1}\sqrt{\pi}\Gamma(n+\Delta+1)}\frac{1}{x(1-x^2)^{\frac{\Delta-1}{2}}}Q_{2n+\Delta+1}^{\Delta-1}(x)
\end{equation}
and the results from the cancellation of the non-analytic part
\begin{equation}
a^{CFT}_{n}c_{n}[\Delta_{n}(\Delta_{n}-3)-\Delta(\Delta-3)]=\lambda
\frac{(-1)^{n}(n+1)\Gamma(n+\Delta+1)\Gamma(n+\Delta+\frac{3}{2})}{\Gamma(\Delta-\frac{1}{2})\Gamma(\frac{1}{2})\pi^3\Gamma(2n+\Delta+\frac{3}{2})}.
\end{equation}
The summation of the analytic part of the higher dimension operators that we need for the equation of motion (\ref{eomsumv}) is then ($x=\sqrt{1-\chi}$)
\begin{eqnarray}
& &\frac{\lambda}{(y_1-y_2)^{2\Delta+2}}\left[\frac{z}{z^2+(x-y_2)^{2}}\right]^{1-\Delta}\frac{(-1)^{\Delta}\Gamma(-\frac{1}{2})}{2^{(\Delta-1)}\pi^4\Gamma(\Delta-\frac{1}{2})}\frac{1}{x(1-x^2)^{\frac{\Delta-1}{2}}}\nonumber\\
&\times &\sum_{n=0}^{\infty}(n+1)(n+\Delta+\frac{1}{2})(2n+\Delta+\frac{3}{2})Q_{2n+\Delta +1}^{\Delta-1}(x)
\label{anasumv}
\end{eqnarray}
In order to sum this series we start with a known sum \cite{Prudnikov:1990:IS} valid for $x>1$ 
\begin{equation}
\sum_{k=m}^{\infty}(\pm1)^{k}(2k+1)Q_{k}^{m}(x)=(\mp 2)^m \Gamma(m+1)(x\pm1)^{-\frac{m}{2}}(x\mp 1)^{-\frac{m}{2}-1}
\end{equation}
To get the sum for $0<x<1$ we need to use that for $-1<x<1$
\begin{equation}
Q_{k}^{n}(x)=\frac{1}{2}e^{-in\pi}(e^{-\frac{in\pi}{2}}Q^{n}_{k}(x+i\epsilon)+e^{\frac{in\pi}{2}}Q^{n}_{k}(x-i\epsilon))
\end{equation}
so that for $0<x<1$ we get
\begin{equation}
\sum_{k=m}^{\infty}(\pm1)^{k}(2k+1)Q_{k}^{m}(x)=(\mp 2)^m \Gamma(m+1)(1\pm x)^{-\frac{m}{2}}(1\mp x)^{-\frac{m}{2}}\frac{1}{x \mp 1}
\end{equation}
We thus get for $0<x<1$
\begin{equation}
\sum_{n=0}^{\infty}(4n+2\Delta-1)Q_{2n+\Delta-1}^{\Delta-1}(x)=(-2)^{\Delta-1}\Gamma(\Delta)(1-x^2)^{-\frac{\Delta-1}{2}}\frac{x}{x^2-1}
\label{anasumstv}
\end{equation}
To get an expression like (\ref{anasumv}) we act on both sides of (\ref{anasumstv}) with $L_{\Delta-1}^{\Delta-1}$ to get
\begin{equation}
\sum_{n=0}^{\infty}(4n+2\Delta+3)(2n+2\Delta+1)(2n+2)Q_{2n+\Delta+1}^{\Delta-1}(x)=4(-2)^{\Delta-1}\Gamma(\Delta+1)\frac{x}{(1-x^2)^{\frac{\Delta+3}{2}}}
\label{anasumvfin}
\end{equation}
Putting everything together we get that (\ref{anasumv}) is just
\begin{equation}
\frac{\lambda}{(y_1-y_2)^{2\Delta+2}}\left[\frac{z}{z^2+(x-y_2)^{2}}\right]^{1-\Delta}\frac{\Gamma(\Delta+1)}{\pi^3 \Gamma(\Delta-\frac{1}{2})\Gamma(\frac{1}{2})}\frac{1}{\chi^{\Delta+1}}
\end{equation}
This is just
\begin{equation}
\frac{\lambda \Gamma(\Delta+1)}{\pi^3 \Gamma(\Delta-\frac{1}{2})\Gamma(\frac{1}{2})}\left(\frac{z}{z^2+(x-y_1)^2}\right)^{\Delta+1}\left(\frac{z}{z^2+(x-y_2)^2}\right)^{2}
\end{equation}
so we indeed have the expected bulk equation of motion.

\section{Scalar system relevant for gravity\label{appendix:scalargravity}}
In this subsection we treat the scalar system relevant for the scalar-gravity interaction. As seen in section \ref{sect:gravity}, the computation of a dimension $\Delta$ primary scalar operator uplifted into the bulk interacting with gravity is related to a scalar system with operators of dimension $\Delta, \Delta+2, d$. This is the system we treat in this appendix, and since the criterion for locality we use is only good for $d\geq 4$, we study the cases $d=4,5$. What we will show is that the coefficients $a_{n}^{CFT}c_{n}$ match the form in (\ref{coefbulk}) and thus also the result from the bulk computation of appendix \ref{appendix:bulk}. This establishes that the sum of the higher dimension operators appearing on the right hand side of the equation of motion is just $\phi_{j}^{(0)}\phi_{k}^{(0)}(x,z)$, and that the sum over the analytic part is
\begin{equation}
\frac{\lambda}{N}\frac{\Gamma(\Delta+2)\Gamma(d)}{\pi^{d}\Gamma(\Delta+2-\frac{d}{2})\Gamma(\frac{d}{2})}\left(\frac{z}{(x-y_1)^2+z^2}\right)^{\Delta+2}\left(\frac{z}{(x-y_1)^2+z^2}\right)^{d}
\end{equation}

\subsection{$d=4$ case}
As explained in appendix \ref{appendix:polecut}, we are looking for coefficients $a_{n}^{CFT}$ that obey
\begin{eqnarray}
& & \tilde{\gamma}\frac{\Gamma(\Delta_{\nu}-1)}{\Gamma(\Delta_{0}^{(\nu)})\Gamma(\Delta_{\nu}-\Delta_{0}^{(\nu)})}F(\Delta_{0}^{(\nu)},2+\Delta_{0}^{(\nu)}-\Delta_{\nu},2,1-\chi)\nonumber\\
& &= -\frac{1}{N}\sum_{n=0}^{\infty} a^{CFT}_{n}c_{n}^{jk}\frac{\Gamma(\Delta_{n}-1)}{\Gamma(\Delta_{0}^{(n)})\Gamma(\Delta_{n}-\Delta_{0}^{(n)})}F(\Delta_{0}^{(n)},2+\Delta_{0}^{(n)}-\Delta_{n},2,1-\chi)\nonumber\\
\label{firstd4}
\end{eqnarray}
In the case at hand
\begin{eqnarray}
& & \Delta_{\nu}=\Delta, \ \ \Delta_{0}^{(\nu)}=\Delta-1,\ \ \Delta_{n}=2n+\Delta+6,\ \ \Delta_{0}^{(n)}=n+\Delta+2 \nonumber\\
& & \tilde{\gamma}=-\frac{\lambda}{N}\frac{\Gamma(\Delta+1)}{2\pi^4 \Gamma(\Delta)}
\end{eqnarray}
So we are looking to write $\tilde{\gamma}F(\Delta-1,1,2,1-\chi)$ as an infinite sum of $F(k+\Delta,-k,2,1-\chi)$ with $k\geq 2$.

Using some hypergeometric identities we get
\begin{equation}
F(\Delta-1,1,2,1-\chi)=\frac{1}{\Delta-2}\left(\frac{1}{(1-\chi)\chi^{\Delta-2}}-\frac{1}{1-\chi}\right)
\end{equation}
On the other hand it is useful to write the higher dimension operator contribution using Jacobi polynomials
\begin{equation}
F(k+\Delta,-k,2,1-\chi)=\frac{(-1)^k \Gamma(k+1)}{\Gamma(k+2)}P_{k}^{(\Delta-2,1)}(1-2\chi)
\end{equation}
So labeling $1-\chi=y$ we see that we need coefficients $\alpha_{k}$ satisfying
\begin{eqnarray}
\frac{1}{\Delta-2}\left(\frac{2^{\Delta-1}}{(1+y)(1-y)^{\Delta-2}}-\frac{2}{1+y}\right)=\sum_{k}\alpha_{k}P_{k}^{(\Delta-2,1)}(y)
\end{eqnarray}
Using the orthogonality condition for the Jacobi polynomials
\begin{equation}
\int_{-1}^{1}P_{k}^{(\Delta-2,1)}(y)P_{l}^{(\Delta-2,1)}(y)(1-y)^{\Delta-2}(1+y)=\delta_{kl}\frac{2^{\Delta}\Gamma(k+\Delta-1)\Gamma(k+2)}{\Gamma(k+1)\Gamma(k+\Delta)(2k+\Delta)}
\end{equation}
and the integrals
\begin{eqnarray}
& & \int_{-1}^{1}(1-y)^{\Delta-2} P_{k}^{(\Delta-2,1)}(y)=(-1)^k2^{\Delta-1}\frac{\Gamma(k+\Delta-1)}{\Gamma(k+\Delta)}\nonumber\\
& & \int_{-1}^{1}P_{k}^{(\Delta-2,1)}(y)=2\frac{\Gamma(k+\Delta-1)}{\Gamma(k+\Delta)}\left[\frac{\Gamma(k+\Delta-1)}{\Gamma(k+2)\Gamma(\Delta-2)}+(-1)^{k}\right]\nonumber
\end{eqnarray}
we find
\begin{equation}
\alpha_{k}=\frac{\Gamma(k+\Delta-1)\Gamma(k+1)(2k+\Delta)}{(\Gamma(k+2))^2\Gamma(\Delta-1)}
\end{equation}
So we can write
\begin{eqnarray}
F(\Delta-1,1,2,1-\chi)=\sum_{k=0}^{\infty}(-1)^k\frac{\Gamma(k+\Delta-1)(2k+\Delta)}{\Gamma(\Delta-1)\Gamma(k+2)}F(k+\Delta,-k,2,1-\chi)\nonumber \\
\label{basumd4}
\end{eqnarray}

We need however to find a formula without the $k=0,1$ terms which are missing from the contribution of the available higher-dimension operators. To do this we act on (\ref{basumd4}) with a differential operator that eliminates these terms. 
The hypergeometric function solves the equation
\begin{equation}
[x(1-x)\frac{d^2}{d^2 x}+(c-(a+b+1)x)\frac{d}{dx}-ab]F(a,b,c,x)=0
\end{equation}
Note that  all the hypergeometric functions appearing in (\ref{basumd4}) have the same values $c=2$ and $a+b+1=\Delta+1$. We define ($z=1-\chi$)
\begin{equation}
L_{G,2}^{\Delta+1}=z(1-z)\frac{d^2}{d^2 z}+(2-(\Delta+1)z)\frac{d}{dz}
\end{equation}
The needed operator is then
\begin{equation}
L_{G,2}^{\Delta+1}(L_{G,2}^{\Delta+1}+\Delta+1)
\end{equation}
We now act with this operator on both side of  (\ref{basumd4}).  Writing $k=n+2$ we get the formal sum
\begin{eqnarray}
& & F(\Delta-1,1,2,1-\chi)=\sum_{n=0}^{\infty}(-1)^n\frac{\Gamma(n+\Delta+1)(2n+\Delta+4)}{2\Gamma(n+4)\Gamma(\Delta+1)}\\
& &\times (n+2)(n+3)(n+\Delta+2)(n+\Delta+3)F(n+\Delta+2,-n-2,2,1-\chi)\nonumber
\end{eqnarray}
We can now read off the coefficients that we want in equation (\ref{firstd4}).
\begin{equation}
a^{CFT}_{n}c_{n}=\frac{\lambda(-1)^n}{4\pi^4 \Gamma(\Delta)}\frac{\Gamma(n+\Delta+1)\Gamma(n+\Delta+4)\Gamma(n+3)}{\Gamma(n+1)\Gamma(2n+\Delta+4)}
\end{equation}
After a bit of algebra this exactly matches the result (\ref{coefbulk}).

\subsection{$d=5$ case}
We start with a scalar 3-point function of  dimensions $(\Delta, \Delta +2, 5)$. The discontinuity given by (\ref{discont}) is
\begin{equation}
\tilde{\gamma}^{d=5}_{\Delta,\Delta+2,5}\frac{1}{(y_1-y_2)^{2\Delta+4}}\left[\frac{z}{z^2 +(x-y_2)^2}\right]^{(3-\Delta)}\frac{1}{(1-\chi)^{\frac{3}{2}}}\frac{1}{\Gamma(\frac{3}{2})\Gamma(-\frac{1}{2})}\frac{1}{\chi^{\Delta-3}}
\end{equation}
The discontinuity coming from higher-dimension operators of dimension $\Delta_{n}=2n+\Delta+7$ is
\begin{equation}
\frac{c_{n}}{(1-\chi)^{\frac{3}{2}}}\frac{\Gamma(2n+\Delta+5\frac{1}{2})}{\Gamma(n+\Delta+2)\Gamma(n+5)\Gamma(-\frac{1}{2})}F(n+\Delta+\frac{1}{2},-n-4,-\frac{1}{2},1-\chi)
\end{equation}
We can use the representation of the hypergeometric function in terms of Jacobi polynomials $P^{(\alpha,\beta)}_{k}$ to write
\begin{equation}
F(n+\Delta+\frac{1}{2},-n-4,-\frac{1}{2},1-\chi)=(-1)^n\frac{\Gamma(n+5)\Gamma(-\frac{1}{2})}{\Gamma(n+3\frac{1}{2})}P^{(\Delta-3,-\frac{3}{2})}_{n+4}(1-2\chi).
\end{equation}
The discontinuity from the higher-dimension operators is then
\begin{equation}
(-1)^n \frac{c_{n}}{(1-\chi)^{\frac{3}{2}}} \frac{\Gamma(2n+\Delta+5\frac{1}{2})}{\Gamma(n+\Delta+2)\Gamma(n+3\frac{1}{2})}P^{(\Delta-3,-\frac{3}{2})}_{n+4}(1-2\chi)
\end{equation}
Apparently we need to find a formula to write
\begin{equation}
\frac{1}{\chi^{\Delta-3}}=\sum_{k} \alpha_{k} P_{k}^{(\Delta-3,-\frac{3}{2})}(1-2\chi)
\label{sumd5}
\end{equation}
In a more convenient form we label $1-2\chi=y$ and want to write
\begin{equation}
\frac{2^{\Delta-3}}{(1-y)^{\Delta-3}}=\sum_{k}\alpha_{k} P_{k}^{(\Delta-3,-\frac{3}{2})}(y)
\end{equation}
To find the $\alpha_{k}$ we use orthogonality of the Jacobi polynomials $P_{n}^{\alpha,\beta}$
\begin{eqnarray}
& &\int_{-1}^{1}dx P^{(\Delta-3,-\frac{3}{2})}_{n}(x)P^{(\Delta-3,-\frac{3}{2})}_{k}(x)(1-x)^{\Delta-3}(1+x)^{-\frac{3}{2}}=\nonumber\\
& &\delta_{nk}\frac{\Gamma(n+\Delta-2)\Gamma(n-\frac{1}{2})2^{(\Delta-3\frac{1}{2})}}{\Gamma(n+1)\Gamma(n+\Delta-3\frac{1}{2})(2n+\Delta-3\frac{1}{2})}
\end{eqnarray}
and the integral
\begin{equation}
\int_{-1}^{1}dx(x+1)^{\sigma}(1-x)^{\beta-1}P_{n}^{(\rho,\sigma)}(x)=2^{\beta+\sigma}\frac{\Gamma(\rho-\beta+1+n)}{\Gamma(n+1)\Gamma(\rho-\beta+1)}B(\beta,\sigma+n+1)
\end{equation}
One gets
\begin{equation}
\alpha_{k}=\frac{\Gamma(k+\Delta-3)\Gamma(k+\Delta-3\frac{1}{2})(2k+\Delta-3\frac{1}{2})}{\Gamma(k+\Delta-2)\Gamma(k+\frac{1}{2})\Gamma(\Delta-3)}
\end{equation}
However we do not have from the higher dimension operators the first four Jacobi polynomials with $0\leq k \leq 3$. We can get an equation of the sort we want  with the usual trick of acting on both sides of (\ref{sumd5}) with an operator of which all terms are eigenfunctions.  Note that
\begin{equation}
\frac{1}{\chi^{\Delta-3}}=F(\Delta-3,-\frac{1}{2}, -\frac{1}{2}, 1-\chi)
\end{equation}
and as noted above the higher dimension operators contribute 
\begin{equation}
F(n+\Delta+\frac{1}{2},-n-4,-\frac{1}{2},1-\chi)
\end{equation}
So all terms are of the form $F(a,b,c,x)$ with the same $c=-\frac{1}{2}$ and the same $a+b+1=\Delta-\frac{5}{2}$.
Since the hypergeometric function solves the equation
\begin{equation}
[x(1-x)\frac{d^2}{d^2 x}+(c-(a+b+1)x)\frac{d}{dx}-ab]F(a,b,c,x)=0
\end{equation}
we see that we can define our needed operator
\begin{equation}
L_{G,-\frac{1}{2}}^{\Delta-2\frac{1}{2}}=x(1-x)\frac{d^2}{d^2 x}+(-\frac{1}{2}-(\Delta-\frac{5}{2})x)\frac{d}{dx}
\end{equation}
To eliminate the first four terms we need to act with (converting the Jacobi polynomials back to hypergeometric functions and labeling by $x$ the argument of the hypergeometric function)
\begin{equation}
L_{G,-\frac{1}{2}}^{\Delta-2\frac{1}{2}}[L_{G,-\frac{1}{2}}^{\Delta-2\frac{1}{2}}+(\Delta-2\frac{1}{2})][L_{G,-\frac{1}{2}}^{\Delta-2\frac{1}{2}}+2(\Delta-\frac{3}{2})][L_{G,-\frac{1}{2}}^{\Delta-2\frac{1}{2}}+3(\Delta-\frac{1}{2})]
\end{equation}
We act on (\ref{sumd5}) with this operator.  On the left we get a multiplicative factor $-\frac{15}{16}\frac{\Gamma(\Delta+1)}{\Gamma(\Delta-3)}$.  On the right we eliminate the first four terms and each other term gets multiplied by
$\frac{\Gamma(\Delta+k+\frac{1}{2})\Gamma(k+1)}{\Gamma(\Delta+k-\frac{7}{2})\Gamma(k-3)}$.
Thus formally we get the desired formula
\begin{eqnarray}
& &\frac{1}{\chi^{\Delta-3}}=\sum_{n=0}^{\infty} \alpha_{n} P_{n+4}^{(\Delta-3,-\frac{3}{2})}(1-2\chi)\\
& &\alpha_{n}=-\frac{16}{15\Gamma(\Delta+1)}\frac{\Gamma(n+\Delta+1)\Gamma(n+5)\Gamma(n+\Delta+4\frac{1}{2})}{\Gamma(n+\Delta+2)\Gamma(n+1)\Gamma(n+4\frac{1}{2})}
\end{eqnarray}
Using from (\ref{corgijk})
\begin{equation}
\tilde{\gamma}^{d=5}_{\Delta,\Delta+2,5}=-\frac{\lambda}{N}\frac{1}{4\pi^5}\frac{\Gamma(\frac{3}{2})\Gamma(\frac{7}{2})\Gamma(\Delta+1)}{\Gamma(\Delta-\frac{1}{2})\Gamma(\frac{5}{2})}
\end{equation}
we arrive at an expression for the coefficient
\begin{equation}
a_{n}^{CFT}c_{n}=\frac{\lambda(-1)^n}{\pi^5 \Gamma(\Delta-\frac{1}{2})\Gamma(\frac{5}{2})}\frac{\Gamma(n+5)\Gamma(n+\Delta+4\frac{1}{2})\Gamma(n+\Delta+2)}{\Gamma(n+1)\Gamma(2n+\Delta+4\frac{1}{2})}\frac{1}{(2n+2\Delta+2)(2n+7)}
\end{equation}
which exactly agrees with the general result (\ref{coefbulk}).

\section{Metric - Weyl correlator\label{appendix:hC}}
Here we derive equation (\ref{ttbulk}).
We start with the expression for the $\langle T_{\mu \nu}T_{\alpha \beta}\rangle $ correlator in appendix A of \cite{Erdmenger:1996yc}.
From all the terms appearing in this expression the only one that will contribute to a $\langle T_{\mu\nu}C_{\alpha \beta \gamma \delta}\rangle $ correlator with $\alpha,\beta,\gamma,\delta$ all distinct is
\begin{equation}
\langle T_{\mu\nu}(x)T_{\alpha\beta}(y)\rangle =\frac{C_{T}}{8(d-2)^2d(d+1)}(\nabla^2)^2 (\eta_{\mu \alpha}\eta_{\nu \beta}+\eta_{\mu \beta}\eta_{\nu \alpha})\frac{1}{(x-y)^{2d-4}}+\cdots
\label{ttcft}
\end{equation}
To compute $C_{T}$ we use results from \cite{Liu:1998bu}. For a canonical normalization of the fluctuations of the metric (that is, the original fluctuations are multiplied by $\kappa$) as used in 
\cite{Liu:1998bu}, one gets
\begin{equation}
C_{T}=\frac{2d(d+1)\Gamma(d)}{\pi^{\frac{d}{2}}(d-1)\Gamma(\frac{d}{2})}
\end{equation}
Putting this in (\ref{ttcft}) one gets
\begin{equation}
\langle T_{\mu\nu}(x)T_{\alpha\beta}(y)\rangle =\frac{d\Gamma(d)}{\pi^{\frac{d}{2}}\Gamma(\frac{d}{2})}(\eta_{\mu \alpha}\eta_{\nu \beta}+\eta_{\mu \beta}\eta_{\nu \alpha})\frac{1}{(x-y)^{2d}}+\cdots
\label{ttcft1}
\end{equation}
Now  $z^2 T_{\mu \nu}$ smears like a scalar of dimension $d$, but we need to take into account the normalization factor $\frac{1}{2\Delta-d}=\frac{1}{d}$, so overall we find that the relevant term for a bulk-boundary correlator is
\begin{equation}
\langle z^2h^{(0)}_{\mu\nu}(x,z)T_{\alpha\beta}(y)\rangle =\frac{\Gamma(d)}{\pi^{\frac{d}{2}}\Gamma(\frac{d}{2})}(\eta_{\mu \alpha}\eta_{\nu \beta}+\eta_{\mu \beta}\eta_{\nu \alpha})\left(\frac{z}{(x-y)^{2}+z^2}\right)^d+\cdots
\label{ttcft2}
\end{equation}
Thus we arrive at
\begin{eqnarray}
& &\langle z^2h^{(0)}_{\mu \nu}(x,z) C_{\alpha \beta \gamma \delta}(y_2)\rangle =\frac{\Gamma(d)}{\pi^{\frac{d}{2}}\Gamma(\frac{d}{2})}[(\eta_{\mu \beta}\eta_{\nu \delta}+\eta_{\nu \beta}\eta_{\mu \delta})\partial^{y_2}_{\alpha}\partial^{y_2}_{\gamma} -(\gamma \leftrightarrow \delta)-\nonumber\\
& &(\beta \leftrightarrow \alpha)+(\gamma \leftrightarrow \delta \ \ \beta \leftrightarrow \alpha)]
\left(\frac{z}{(x-y_2)^2+z^2}\right)^{d}
\label{ttbulk1}
\end{eqnarray}

\providecommand{\href}[2]{#2}\begingroup\raggedright\endgroup

\end{document}